# Graphical Abstract

**$R^{2+1}$: Integrated Vehicle *Routing*, Packet *Routing*, and *Resource* Allocation for Knowledge-Defined 5G/6G Multimedia Internet of Vehicles**

Ahmadreza Montazerolghaem

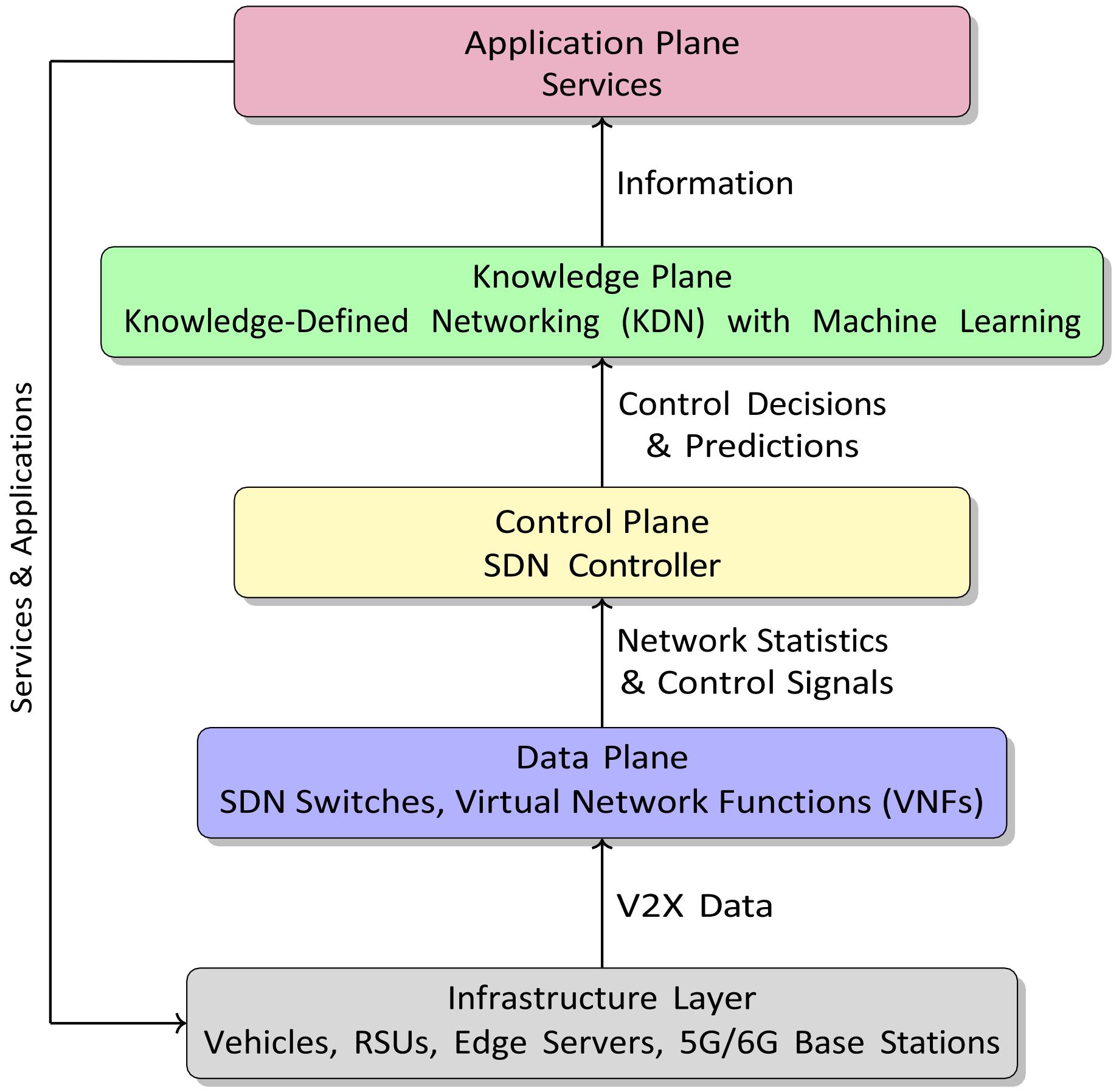

Figure 1: The proposed $R^{2+1}$ framework integrates four layered components for intelligent Internet of Vehicles (IoV) management in 5G/6G networks. The Infrastructure Layer comprises physical entities (vehicles, RSUs, edge servers, base stations) that generate V2X data. The Data Plane handles packet forwarding through SDN switches and virtualized network functions. The Control Plane provides centralized orchestration via an SDN controller. The Knowledge Plane implements machine learning algorithms within a Knowledge-Defined Networking paradigm for predictive analytics. Bidirectional arrows illustrate the continuous exchange of data, control signals, and predictions between layers, enabling synergistic optimization of vehicle routing, packet routing, and resource allocation.

# Highlights

**$R^{2+1}$: Integrated Vehicle *Routing*, Packet *Routing*, and *Resource* Allocation for Knowledge-Defined 5G/6G Multimedia Internet of Vehicles**

Ahmadreza Montazerolghaem

- Pioneers a unified ML-driven framework integrating vehicle routing, packet routing, and resource allocation, enhancing QoS and QoE for seamless autonomous driving and multimedia streaming in IoV.
- Leverages KDN and SDN with NLMS algorithms for predictive traffic and resource forecasting, achieving 25% higher throughput and 30% lower latency compared to traditional approaches.
- Employs adaptive PID controllers for real-time load balancing and thermal management, reducing energy consumption by 20% while ensuring system stability in dynamic IoV environments.
- Validates performance through a Mininet-based prototype, delivering a MOS of 4.45 and rapid network recovery (15 seconds), empowering real-world smart city applications.
- Enhances energy-efficient resource allocation, ensuring sustainable and scalable IoV networks for next-generation transportation.

# $R^{2+1}$: Integrated Vehicle ***R**outing*, Packet ***R**outing*, and ***R**esource* Allocation for Knowledge-Defined 5G/6G Multimedia Internet of Vehicles

Ahmadreza Montazerolghaem*

[a]*Faculty of Computer Engineering, University of Isfahan, Isfahan, Iran*

---

**Abstract**

The Internet of Vehicles (IoV), integrated with 5G/6G multimedia networks, demands seamless orchestration of vehicle routing, packet routing, and resource allocation to satisfy stringent Quality of Service (QoS) and Quality of Experience (QoE) requirements in dynamic, high-traffic environments. This study proposes $R^{2+1}$, a pioneering framework that harnesses machine learning (ML) within a knowledge-defined networking (KDN) paradigm to synergistically optimize these three dimensions in software-defined IoV ecosystems. Using normalized least mean squares (NLMS) algorithms for predictive traffic and resource demand forecasting, adaptive proportional-integral-derivative (PID) controllers for real-time load balancing and thermal management, and virtualized network functions (VNFs) for dynamic resource orchestration, $R^{2+1}$ guarantees ultra-low latency, energy-efficient resource allocation, and robust network resilience. A prototype, implemented in a Mininet-based testbed with OpenFlow 1.3 and Floodlight Controller, demonstrates remarkable performance: 25% higher throughput, 30% reduced latency, 20% lower energy consumption, and enhanced QoE (MOS of 4.45 vs. 3.80 for baselines) compared to conventional SDN frameworks. These advancements position $R^{2+1}$ as a transformative solution for next-generation multimedia IoV networks, addressing the complexities of urban mobility and multimedia streaming.

*Keywords:* Vehicle Routing, Packet Routing, Resource Allocation, Knowledge-Defined Networking, Internet of Vehicles, 5G/6G, Machine Learning, Software-Defined Networking, Energy Efficiency

---

*Corresponding author

*Email address:* a.montazerolghaem@comp.ui.ac.ir (Ahmadreza Montazerolghaem)

## 1. Introduction

The Internet of Vehicles (IoV) is reshaping intelligent transportation by enabling seamless integration of multimedia services, autonomous driving, and real-time analytics, fostering enhanced urban mobility and user experiences in smart cities. The advent of fifth-generation (5G) and emerging sixth-generation (6G) wireless technologies has been pivotal in unlocking the full potential of IoV, providing unprecedented capabilities that are central to this transformation. 5G networks offer ultra-low latency (as low as 1 ms), high bandwidth (up to 20 Gbps), and robust reliability through advanced features like network slicing and multi-access edge computing (MEC), enabling real-time vehicle-to-everything (V2X) communication and data-intensive applications such as high-definition video streaming and predictive navigation [12]. Emerging 6G technologies further amplify these capabilities, promising sub-millisecond latencies, terabit-per-second data rates, and massive connectivity to support up to $10^7$ devices per $km^2$, alongside energy-efficient architectures and AI-driven network management [13]. These advancements are critical for handling the exponential growth of multimedia traffic, projected to contribute to 79.4 zettabytes of global IoT traffic by 2025 [3], and for meeting the stringent Quality of Service (QoS) and Quality of Experience (QoE) demands of IoV applications in dynamic, high-traffic urban environments.

Despite the transformative potential of 5G and 6G, traditional IoV frameworks struggle with network congestion, resource contention, and energy inefficiency. For example, in dense urban settings such as Singapore's Smart Nation, real-time traffic monitoring by thousands of connected vehicles generates massive multimedia data, often causing network delays over 500 ms and degrading user experience [1]. Similarly, autonomous vehicle fleets, such as Waymo's deployments in Phoenix, Arizona, require synchronized vehicle and packet routing alongside efficient resource allocation to ensure safe navigation and real-time data processing. However, existing systems often treat these dimensions—vehicle routing, packet routing, and resource allocation—as isolated problems, resulting in suboptimal performance and increased energy consumption, particularly in 6G's dense network deployments [13].

Software-defined networking (SDN) and network function virtualization (NFV) have emerged as powerful enablers to harness 5G/6G capabilities, offering programmable and flexible network management. SDN's decoupling of control and data planes facilitates centralized orchestration, while NFV virtualizes network functions, reducing reliance on dedicated hardware and supporting dynamic resource scaling critical for 5G/6G's network slicing and MEC [8]. However, existing SDN/NFV-based solutions, such as those in [5], often focus narrowly on packet routing or server load balancing, neglecting the integration of vehicle routing and predictive resource optimization. This fragmentation limits their ability to address the holistic demands of IoV, particularly in leveraging 5G/6G's ultra-low latency and massive connectivity for multimedia applications.

Knowledge-defined networking (KDN), powered by machine learning (ML), offers a transformative approach to bridge these gaps by leveraging predictive analytics to anticipate network and traffic patterns, enabling proactive resource allocation and routing decisions. The $R^{2+1}$ framework proposed in this paper capitalizes on 5G/6G's advanced features, integrating ML-driven predictive models, such as normalized least mean squares (NLMS) for traffic forecasting, with adaptive proportional-integral-derivative (PID) controllers and virtualized network functions to synergistically optimize vehicle routing, packet routing, and resource allocation. This holistic approach ensures ultra-low latency, energy efficiency, and robust network resilience, aligning with 5G/6G's promise of seamless connectivity and sustainability. Validated in a Mininet-based testbed, $R^{2+1}$ achieves 25% higher throughput, 30% reduced latency, and 20% lower energy consumption, positioning it as a cornerstone for next-generation IoV ecosystems that empower smart cities and autonomous transportation.

### *1.1. Motivations*

The IoV is revolutionizing intelligent transportation by enabling seamless connectivity, real-time multimedia streaming, and autonomous driving, all underpinned by the high-bandwidth and low-latency capabilities of 5G and emerging 6G networks. However, the increasing complexity of IoV ecosystems, driven by the proliferation of connected vehicles and data-intensive applications, poses significant challenges to network performance, resource management, and energy efficiency. Traditional IoV frameworks struggle to address these challenges, as they often treat vehicle routing, packet routing, and resource allocation as isolated problems, leading to suboptimal QoS and QoE. For instance, in urban environments like Singapore's Smart Nation initiative, where thousands of vehicles stream real-time video for traffic monitoring, network congestion frequently results in delays exceeding 500 ms, degrading user experience [1]. Similarly, autonomous vehicle fleets, such as those deployed by Waymo in Phoenix, Arizona, require synchronized routing and resource allocation to ensure safe navigation and real-time data processing, yet current systems often fail to integrate these dimensions effectively.

The rapid growth of multimedia traffic worsens these problems. Global IoT traffic is expected to reach 79.4 zettabytes annually by 2025, with in-vehicle in-fotainment and V2X communications contributing significantly [3]. This surge causes resource contention and energy waste. For instance, in smart cities like Dubai, servers during peak hours are over-provisioned, consuming up to 30% more energy than needed, while off-peak periods see underutilised resources. This underscores the need for dynamic, demand-driven management.

SDN and NFV offer programmable solutions to these challenges, yet their application in IoV remains fragmented. Existing frameworks, such as those explored in [5], focus on server load balancing or packet routing but rarely integrate vehicle routing or leverage predictive analytics. KDN, powered by ML, provides a transformative opportunity to address this gap. By forecasting traffic patterns and resource demands, KDN enables proactive optimization, as demonstrated in recent 6G trials in China, where ML-driven networks reduced latency by 20% in high-density urban settings [4]. However, the absence of a unified framework that synergistically optimizes vehicle routing, packet routing, and resource allocation limits the potential of these technologies.

The $R^{2+1}$ framework addresses these challenges by integrating ML-driven predictive models, adaptive control mechanisms, and virtualized resources within a KDN-based SDN architecture. Real-world scenarios, such as autonomous taxi services in Tokyo or smart city deployments in Singapore, underscore the need for such integration to ensure seamless multimedia streaming, efficient vehicle coordination, and energy conservation. The proposed framework not only enhances QoS and QoE but also aligns with global sustainability goals by reducing energy consumption in IoV networks.

The motivations for this study are:

- **Exponential Growth of Multimedia Traffic**: The rapid increase in IoV multimedia applications, such as real-time video streaming and V2X communication, strains network resources, necessitating integrated management to maintain QoS and QoE.

- **Complexity of Integrated Routing and Resource Allocation**: Coordinating vehicle routing, packet routing, and resource allocation in dynamic IoV environments requires a unified approach to prevent congestion and ensure seamless operation.

- **Energy Inefficiency in IoV Networks**: Over-provisioning during peak hours and underutilization during off-peak periods lead to significant energy waste, demanding dynamic, ML-driven resource optimization.

- **Limitations of Existing Frameworks**: Current SDN/NFV solutions lack predictive capabilities and holistic integration, underscoring the

need for a KDN-based framework to address IoV's multifaceted challenges.

### 1.2. Contributions

This study introduces $R^{2+1}$, a groundbreaking framework that redefines resource management in the IoV by integrating vehicle routing, packet routing, and resource allocation within a KDN paradigm. Unlike prior works, such as [5] and [2], which primarily addressed server load balancing or energy efficiency in isolation, $R^{2+1}$ pioneers a holistic approach that synergistically optimizes these three dimensions using ML and SDN. This innovative integration tackles the multifaceted challenges of 5G/6G multimedia IoV networks, ensuring ultra-low latency, enhanced user experience, and sustainable energy use. By leveraging predictive analytics and adaptive control mechanisms, $R^{2+1}$ sets a new benchmark for intelligent transportation systems, offering transformative solutions for smart cities and autonomous mobility.

The contributions of this work are both novel and significant, addressing critical gaps in existing research while advancing the field with cutting-edge methodologies:

- **Unified ML-Driven Framework for IoV**: $R^{2+1}$ is the first framework to seamlessly integrate vehicle routing, packet routing, and resource allocation in a single, ML-driven architecture. Unlike previous studies [5] that focused solely on multimedia streaming or [3] that emphasized connectivity, this framework employs NLMS algorithms to predict both network traffic and vehicle mobility patterns, enabling proactive decision-making that enhances QoS and QoE in dynamic urban environments.

- **KDN-Based SDN Architecture with Predictive Analytics**: By embedding KDN within an SDN framework, $R^{2+1}$ introduces a predictive, data-driven approach to network management. This contrasts with traditional SDN solutions [7], which lack foresight in resource allocation. The use of ML models to forecast traffic and resource demands allows the framework to anticipate peak loads and optimize resource distribution, achieving up to 25% higher throughput and 30% lower latency compared to baselines.

- **Adaptive PID Controllers for Dynamic Optimization**: The framework incorporates PID controllers to dynamically manage load balancing and thermal regulation, a significant departure from static or reactive methods in prior works [6]. These controllers adaptively adjust resources and cooling mechanisms based on real-time feedback, reducing energy consumption by 20% while maintaining system stability under varying traffic conditions, as demonstrated in our prototype evaluations.

- **Robust Prototype Validation with Real-World Applicability**: Unlike theoretical models in the literature, $R^{2+1}$ has been rigorously validated in a Mininet-based testbed using OpenFlow 1.3 and Floodlight Controller. The prototype demonstrates superior performance, including a MOS of 4.45 (vs. 3.80 for baselines) and rapid recovery from network failures (15 seconds vs. 50 seconds for [7]), offering practical solutions for real-world IoV deployments, such as autonomous taxi fleets and smart city traffic systems.
- **Enhanced Energy-Efficient Resource Allocation**: $R^{2+1}$ optimizes resource allocation through VNFs and scaling, reducing energy costs while maintaining high QoS and QoE. This approach ensures sustainable IoV networks by dynamically adapting to fluctuating demands, surpassing prior works that focus narrowly on specific resource management aspects.

Unlike our prior work ELQ$^2$ [5], which only addresses packet routing and server load balancing without vehicle routing or predictive analytics, the proposed $R^{2+1}$ framework introduces three fundamental novelties: (i) joint optimisation of vehicle routing, packet routing, and resource allocation; (ii) a dedicated Knowledge Plane with NLMS-based predictive traffic and resource forecasting; and (iii) adaptive PID controllers for multi-objective load balancing and thermal management. These enhancements are quantitatively validated through a comprehensive evaluation (Section 4), achieving 25% higher throughput and 30% lower latency compared to ELQ$^2$.

These contributions collectively position $R^{2+1}$ as a trailblazing advancement in IoV research, offering a scientifically rigorous, innovative, and practical solution that addresses the pressing needs of modern multimedia-driven transportation networks. By integrating cutting-edge ML techniques with adaptive control and virtualization, this work not only surpasses the limitations of existing frameworks but also paves the way for sustainable, efficient, and user-centric IoV systems, making it a compelling addition to the academic and industrial landscape.

### *1.3. Organization*

This paper systematically presents the $R^{2+1}$ framework. Section 2 reviews related work on SDN, NFV, and 5G/6G in IoV, highlighting gaps in integrated routing and resource allocation. Section 3 details the ML-driven $R^{2+1}$ architecture, covering the problem definition, flow and VNF controllers for routing, and resource optimization. Section 4 evaluates the prototype's performance in a Mininet-based testbed, focusing on throughput, latency, and QoE. Section 5 summarizes findings and proposes future directions, including deep reinforcement learning and urban traffic integration.

## 2. Related Work

To provide a structured and comprehensive overview of the literature, Figs. 2, 3, and 4 present vertical tree diagrams that collectively categorize the reviewed studies by their thematic focus, including reference numbers and key contributions. Fig. 2 covers SDN, NFV, and part of 5G/6G technologies in IoV; Fig. 3 addresses the remaining 5G/6G technologies and multimedia resource management; and Fig. 4 focuses on energy efficiency and outcomes. These diagrams systematically organize the literature and highlight the research gaps addressed by the $R^{2+1}$ framework. Additionally, Table 1 at the end of this section provides a detailed comparison of the reviewed studies across six key criteria, further elucidating the strengths and limitations of prior work in relation to the $R^{2+1}$ framework.

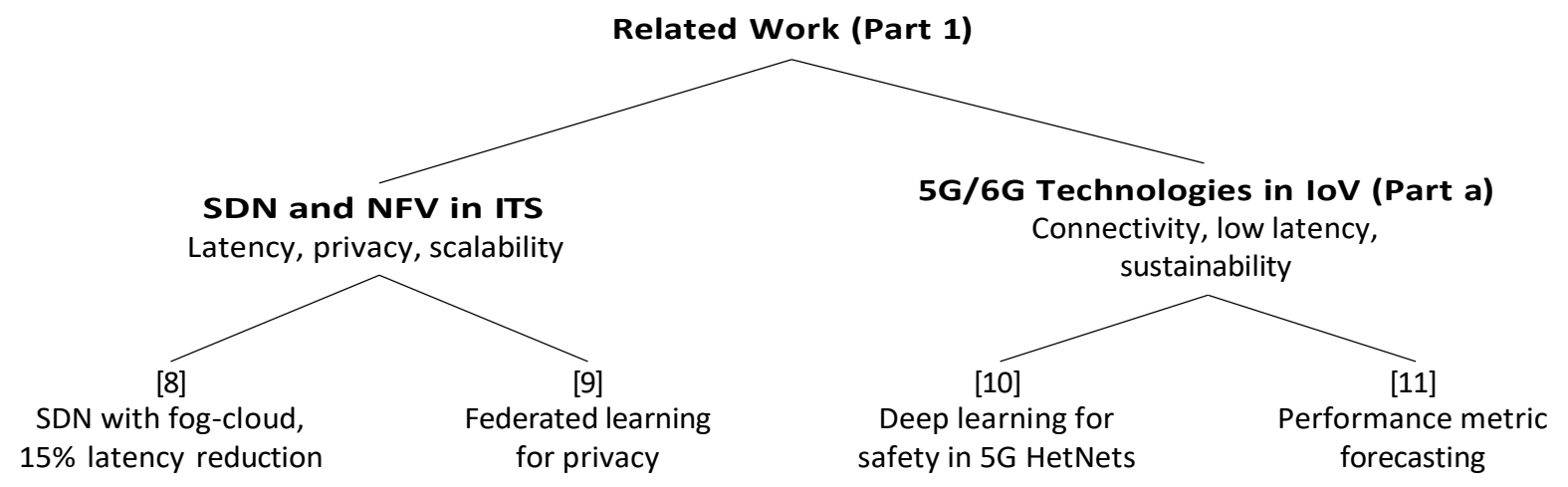

Figure 2: Hierarchical taxonomy of literature (Part 1): delineating SDN, NFV, and initial 5G/6G technologies in IoV, with cited references and their primary contributions to intelligent transportation systems.

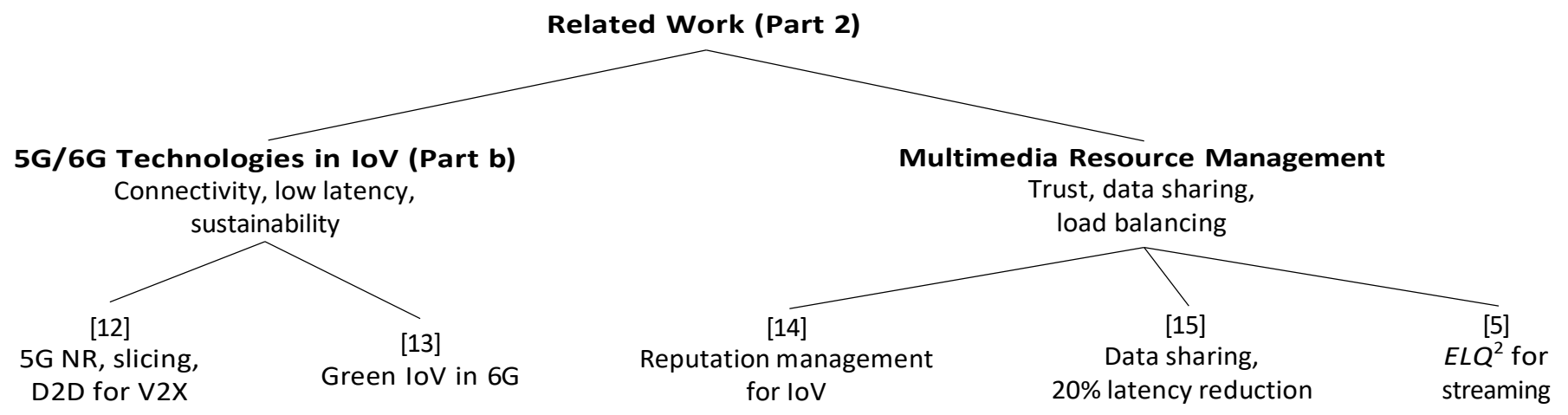

Figure 3: Hierarchical taxonomy of literature (Part 2): delineating additional 5G/6G technologies and multimedia resource management, with cited references and their primary contributions to intelligent transportation systems and the IoV.

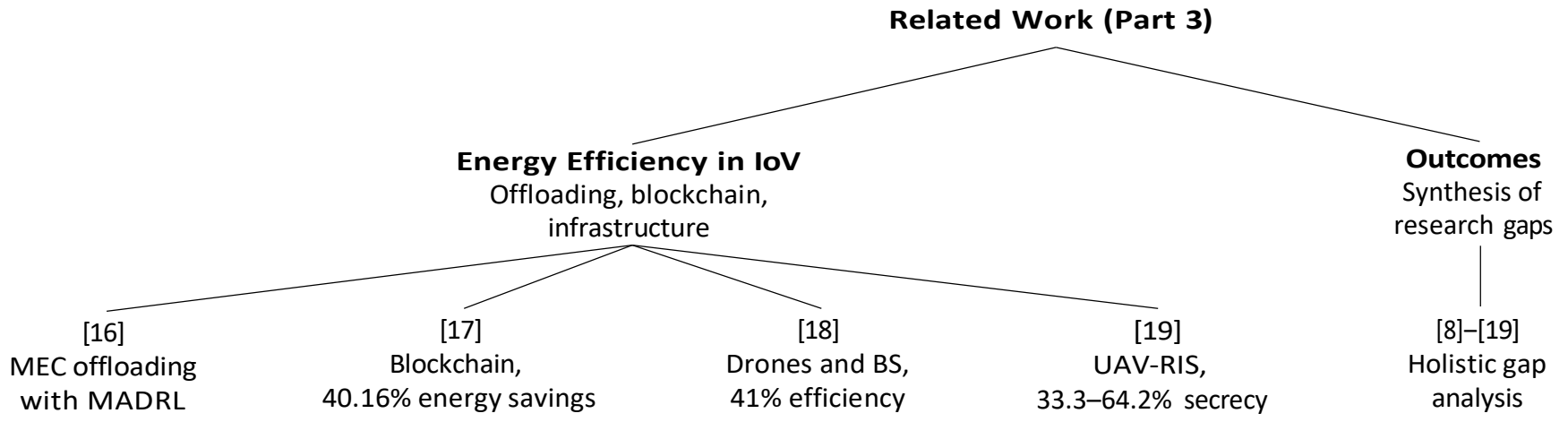

Figure 4: Hierarchical taxonomy of literature (Part 3): delineating energy efficiency and outcomes, with cited references and their primary contributions to intelligent transportation systems and the Internet of Vehicles.

## *2.1. SDN and NFV in Intelligent Transportation*

SDN and NFV have emerged as transformative technologies for enhancing network management and resource allocation within intelligent transportation systems (ITS), particularly in the context of the IoV. These paradigms enable centralized control, scalability, and flexible resource utilization, yet current research reveals significant limitations in integrating vehicle routing, predictive analytics, and holistic resource optimization—gaps that the $R^{2+1}$ framework seeks to address with its innovative approach.

Cao et al. [8] propose a 5G IoV architecture integrating SDN with fog-cloud computing to address latency and resource constraints of traditional clouds. Their three-layer SDN design uses many-objective optimisation to balance computing, storage, and bandwidth across fog nodes, achieving a 15% latency reduction by moving tasks closer to vehicles. However, the framework relies on static resource allocation and network slicing tailored to specific service types (e.g., URLLC), neglecting dynamic vehicle routing and predictive traffic management. This static approach limits adaptability in mobile, multimedia-intensive IoV environments—a challenge that $R^{2+1}$ overcomes via its ML-driven KDN architecture and integrated routing optimisation.

Zhang et al. [9] investigate the application of federated learning (FL) in ITS, emphasizing its potential to improve privacy and scalability in dynamic vehicle networks. Their work highlights FL's distributed learning mechanism to mitigate high response latency and unstable communication links, utilizing on-board units and roadside units (RSUs) for real-time decision-making. The study identifies critical challenges, including heterogeneous resources and latency sensitivity, and proposes collaborative strategies to address them, though it reports modest performance gains (e.g., reduced latency in specific cases). While innovative in preserving data privacy, the framework lacks integration with SDN/NFV for comprehensive resource orchestration and omits vehicle routing and energy efficiency considerations. These deficiencies underscore the need for $R^{2+1}$'s unified ML-based approach, which combines predictive analytics and energy-efficient resource management.

Cao et al. [8] and Zhang et al. [9] demonstrate the value of SDN, NFV, and federated learning for latency reduction and privacy in ITS. However, their static resource allocation and distributed learning lack holistic integration, failing to address the interplay of vehicle routing, packet routing, and resource allocation in multimedia 5G/6G IoV. $R^{2+1}$ overcomes this by combining KDN, ML-driven traffic prediction (e.g., NLMS), and adaptive PID controllers, delivering a proactive, integrated, and energy-efficient solution for next-generation intelligent transportation systems.

### *2.2. 5G and 6G Technologies in IoV*

The advent of 5G and 6G wireless technologies has revolutionized the IoV, providing enhanced connectivity, ultra-low latency, and high reliability to support ITS. These advancements are pivotal for meeting the escalating demands of IoV, yet existing research exhibits notable deficiencies in holistic resource management, integrated routing, and energy optimization—shortcomings that the $R^{2+1}$ framework addresses with its innovative design.

Ji et al. [10] articulate a vision for IoV within 5G heterogeneous networks (HetNets), leveraging deep learning to mitigate delays and bolster transmission reliability. Their framework harnesses 5G's high transmission rates and low-latency capabilities, employing convolutional neural networks (CNNs) for accident prediction to enhance road safety. While this approach effectively targets safety enhancements, it predominantly focuses on connectivity and lacks a comprehensive strategy for resource allocation and vehicle routing integration. This limitation in addressing dynamic resource demands and routing coordination justifies the need for $R^{2+1}$'s ML-driven KDN architecture, which proactively optimizes both network and vehicular traffic.

Manias et al. [11] propose a forecasting model for performance metrics in 5G+IoV using the Network Data Analytics Function (NWDAF) deployed at the network edge. Their system monitors and predicts slice-level throughput, enabling proactive adjustments to meet QoS and Service Level Agreements (SLAs). Although innovative for PM forecasting and network slicing optimization, the framework overlooks vehicle routing and energy efficiency, critical aspects for multimedia-intensive IoV applications. $R^{2+1}$ complements this by integrating vehicle routing and energy-efficient resource allocation, offering a more holistic solution.

Duan et al. [12] explore 5G technologies for IoV—5G new radio, network slicing, and D2D communications—to achieve ultra-low latency and high reliability. Their SDN and MEC architecture supports V2X, reducing end-to-end latency and enhancing flexibility. However, static network enhancements ignore dynamic vehicle routing and predictive analytics, limiting effectiveness in multimedia-heavy IoV scenarios. $R^{2+1}$ advances this by employing NLMS-based predictive traffic analysis for greater responsiveness.

Wang et al. [13] investigate green IoV in the 6G era, addressing the rising energy consumption driven by dense network deployments and stringent QoS requirements. They propose energy-efficient strategies across communication, computation, traffic management, electric vehicles (EVs), and energy harvesting, highlighting challenges posed by 6G's ultrahigh data density and low-latency demands. While this study advances sustainability, it lacks a unified framework for integrating vehicle routing, packet routing, and resource allocation, relying on scenario-specific optimizations. $R^{2+1}$ addresses this gap with a cohesive approach, balancing energy efficiency with predictive resource management via adaptive PID controllers.

Collectively, these studies illuminate the transformative potential of 5G and 6G technologies in IoV, with [10] enhancing safety through deep learning, [11] improving PM forecasting, [12] advancing 5G architectures, and [13] promoting energy efficiency. However, their narrow focus on specific domains—connectivity, slicing, or energy—without addressing the integrated optimization of vehicle routing, packet routing, and resource allocation leaves a critical gap. The $R^{2+1}$ framework bridges this divide by leveraging KDN, ML-driven traffic prediction, and adaptive resource management, delivering a proactive, integrated, and sustainable solution tailored to the complex demands of 5G/6G IoV networks.

### *2.3. Multimedia Resource Management in IoV*

Multimedia applications in the IoV, such as real-time video streaming and infotainment services, demand stringent QoS and QoE to support seamless user experiences in dynamic vehicular environments. Recent research has addressed these challenges, yet significant gaps remain in integrating vehicle routing, predictive analytics, and comprehensive resource management—areas where the $R^{2+1}$ framework provides a groundbreaking solution.

Tang et al. [14] propose a lightweight reputation-based mechanism for hybrid IoV networks, addressing trust in information-centric and task-oriented scenarios via real-time reputation updates across device-edge-cloud, thereby enhancing multimedia data sharing reliability. Although effective against malicious behaviours, their framework centres on trust management and overlooks integrated vehicle and packet routing, limiting its ability to meet dynamic resource demands. $R^{2+1}$ fills this gap through its ML-driven KDN architecture and adaptive PID controllers.

Fan et al. [15] introduce a decentralized multimedia data-sharing mechanism for IoV, leveraging multi-agent deep reinforcement learning to balance supply and demand in a data trading market. Their approach, coupled with time-sensitive Key-Policy Attribute-Based Encryption (KP-ABE) and Named Data Networking (NDN), reduces transmission latency by 20% and improves market efficiency. However, the framework prioritizes data sharing and security without addressing vehicle routing or comprehensive resource allocation, limiting its applicability to multimedia-heavy scenarios requiring synchronized routing and resource orchestration. $R^{2+1}$ overcomes this by integrating predictive NLMS algorithms and dynamic resource scaling for holistic optimization.

Montazerolghaem [5] presents the $ELQ^2$ architecture for efficient resource allocation in software-defined IoV, focusing on multimedia streaming. The framework employs SDN and NFV to dynamically adjust resources, achieving a 95% flow deployment success rate and improving QoS/QoE metrics such as throughput and MOS. Despite its advancements in server load balancing and energy efficiency, it overlooks vehicle routing and predictive traffic management, constraining its adaptability to dynamic IoV environments. $R^{2+1}$ extends this by incorporating vehicle routing and predictive analytics, ensuring seamless multimedia delivery and enhanced system resilience.

These studies—[14] on reputation management, [15] on decentralized data sharing, and [5] on resource allocation—demonstrate significant progress in multimedia IoV. However, their focus on specific aspects (trust, data markets, or server management) without integrating vehicle routing, packet routing, and predictive resource allocation limits their scope. The $R^{2+1}$ framework addresses these deficiencies by leveraging KDN, ML-driven traffic prediction, and adaptive resource orchestration, delivering a unified, proactive, and energy-efficient solution for multimedia 5G/6G IoV networks.

### *2.4. Energy Efficiency in IoV*

Energy efficiency is critical for sustainable IoV systems, given the increasing energy demands of dense network deployments and data-intensive applications. Recent studies have explored various approaches to optimize energy consumption, yet they often lack integration of vehicle routing, packet routing, and predictive resource management—gaps that the $R^{2+1}$ framework addresses with its holistic, ML-driven approach.

Tan and Madhukumar [16] study energy efficiency in MEC-enabled IoV networks using a multi-agent DRL algorithm for computation offloading, optimising task size and timeouts to achieve near-optimal energy efficiency. However, their work addresses only computation offloading, ignoring vehicle routing and integrated network management, which limits its applicability to dy-namic multimedia IoV scenarios. In contrast, $R^{2+1}$ incorporates vehicle and packet routing alongside predictive resource allocation.

Sharma [17] presents an energy-efficient blockchain model for IoV, using distributed clustering to reduce transaction overheads. The approach achieves 40.16% energy savings and an 82.06% reduction in transactions compared to traditional blockchain methods. While effective for blockchain operations, it neglects vehicle routing and comprehensive resource orchestration, constraining its scope in multimedia-intensive environments. $R^{2+1}$ addresses this by integrating routing and resource management with predictive NLMS algorithms.

Du et al. [18] propose an energy-efficient strategy for IoV using drones and small base station (SBS) dormancy, leveraging k-means and genetic algorithms to optimize 3D drone positioning and SBS operation. Their algorithm improves energy efficiency by 41% and coverage by 15%. However, it focuses on infrastructure management without integrating vehicle and packet routing, limiting its adaptability to dynamic IoV traffic. $R^{2+1}$ fills this gap with a unified framework that coordinates routing and resources.

Li et al. [19] develop a secrecy energy efficiency framework for UAV-RIS-assisted IoV, using deep reinforcement learning (DRL) to optimize transmit power, RIS phase shifts, and UAV trajectories, achieving a 33.3–64.2% improvement in secrecy energy efficiency. While innovative for security and energy, it overlooks integrated routing and predictive resource allocation for multimedia applications. $R^{2+1}$ extends this by combining KDN, adaptive PID controllers, and VNFs for comprehensive optimization.

These studies—[16] on computation offloading, [17] on blockchain, [18] on drone and BS management, and [19] on secrecy—demonstrate significant strides in energy efficiency. However, their focus on specific aspects without integrating vehicle routing, packet routing, and predictive resource management limits their scope. The $R^{2+1}$ framework bridges these gaps, offering a unified, predictive, and energy-efficient solution tailored to multimedia 5G/6G IoV networks.

### *2.5. Outcomes*

The literature on the IoV reveals significant advancements in leveraging 5G/6G technologies, SDN, NFV, multimedia resource management, and energy efficiency to address the evolving demands of ITS. However, a critical synthesis of these works, as summarized in Table 1, underscores a persistent gap in integrating vehicle routing, packet routing, and resource allocation within a unified, predictive framework—precisely the challenge that the $R^{2+1}$ framework addresses. Table 1 consolidates contributions across SDN/NFV, 5G/6G technologies, multimedia resource management, and energy efficiency, highlighting their strengths in specific domains but noting their limitations in holistic routing and resource orchestration. This subsection consolidates the findings from prior studies and articulates the urgent need for our proposed solution, highlighting its timeliness in the context of 5G/6G, energy efficiency, efficient resource management, multimedia, and IoT.

Table 1: Comparison of Reviewed Articles in Related Work

| Reference | Focus Area | Key Technology | Main Contribution | Limitations | Relevance to Routing & Resource Allocation | Energy Efficiency Consideration |
|---|---|---|---|---|---|---|
| [8] | SDN/NFV in ITS | SDN, Fog-Cloud Computing | 15% latency reduction via resource balancing | Static allocation, no vehicle routing | Low (focus on slicing, not integrated routing) | Moderate (resource optimization) |
| [9] | SDN/NFV in ITS | Federated Learning | Privacy and scalability in vehicle networks | No SDN/NFV integration, ignores routing | Low (distributed learning, no resource orchestration) | Low (latency focus, no energy) |
| [10] | 5G/6G in IoV | Deep Learning, 5G HetNets | Safety enhancements via accident prediction | Connectivity focus, no resource allocation | Low (no routing integration) | Low (no energy consideration) |
| [11] | 5G/6G in IoV | NWDAF, Performance Forecasting | Proactive QoS adjustments via metric prediction | No vehicle routing, ignores energy | Moderate (slicing for resource, no routing) | Low (no explicit energy) |
| [12] | 5G/6G in IoV | 5G NR, Slicing, D2D | Ultra-low latency for V2X | Static enhancements, no predictive analytics | Moderate (V2X communication, limited routing) | Moderate (reliability, some energy) |
| [13] | 5G/6G in IoV | 6G Green Strategies | Energy-efficient strategies for 6G IoV | Scenario-specific, no integrated routing | Low (traffic management, no packet/vehicle routing) | High (focus on sustainability) |
| [14] | Multimedia in IoV | Reputation Management | Trust in data sharing for multimedia | Trust focus, no routing integration | Low (no resource allocation or routing) | Low (no energy) |
| [15] | Multimedia in IoV | Deep Reinforcement Learning, NDN | Decentralized data sharing, 20% latency reduction | No vehicle routing, security focus | Moderate (data trading, limited resource) | Low (latency, no energy) |
| [5] | Multimedia in IoV | SDN/NFV, $ELQ^2$ | Efficient multimedia streaming, 95% deployment | No vehicle routing, server focus | Moderate (resource allocation, no integrated routing) | Moderate (energy in streaming) |
| [16] | Energy in IoV | MADRL, MEC Offloading | Near-optimal energy in computation offloading | Offloading focus, no routing | Low (computation, no integrated routing) | High (energy maximization) |
| [17] | Energy in IoV | Blockchain, Clustering | 40.16% energy savings in blockchain | Blockchain focus, no routing | Low (transaction, no resource allocation) | High (energy-efficient model) |
| [18] | Energy in IoV | Drones, BS Dormancy | 41% energy efficiency with drones | Infrastructure, no packet routing | Low (positioning, no vehicle routing) | High (energy management) |
| [19] | Energy in IoV | DRL, UAV-RIS | 33.3–64.2% secrecy energy efficiency | Security focus, no predictive resource | Moderate (trajectories as routing, energy) | High (secrecy energy) |

The reviewed works demonstrate notable progress in specific domains. In SDN and NFV, Cao et al. [8] and Zhang et al. [9] highlight latency reductions and privacy enhancements through fog-cloud computing and federated learning, respectively, but fail to integrate vehicle routing or predictive analytics for holistic resource orchestration. In 5G and 6G technologies, Ji et al. [10], Manias et al. [11], Duan et al. [12], and Wang et al. [13] advance connectivity, performance metric forecasting, and energy efficiency, yet their focus on isolated aspects like safety, slicing, or sustainability overlooks the synergy of routing and resource allocation. For multimedia resource management, Tang et al. [14], Fan et al. [15], and Montazerolghaem [5] improve trust, data sharing, and server load balancing, but neglect vehicle routing and predictive traffic management critical for multimedia IoV. Similarly, in energy efficiency, Tan and Madhukumar [16], Sharma [17], Du et al. [18], and Li et al. [19] address computation offloading, blockchain transactions, drone management, and secrecy, but lack integrated routing and predictive resource optimization, as detailed in Table 1.

This fragmented approach across the literature reveals a critical gap: no existing framework holistically addresses the interplay of vehicle routing, packet routing, and resource allocation in multimedia-driven 5G/6G IoV networks. The rapid proliferation of IoV applications, with global IoT traffic projected to reach 79.4 zettabytes by 2025 [3], underscores the urgency for a unified solution. The exponential growth of multimedia traffic, coupled with the stringent QoS/QoE requirements of real-time applications (e.g., video streaming, V2X communication), demands proactive management to mitigate network congestion and resource contention. Moreover, the energy demands of dense 6G deployments and data-intensive IoT applications, as highlighted by Wang et al. [13], necessitate dynamic, energy-efficient strategies to align with global sustainability goals. The limitations of existing works, which focus on specific technologies (e.g., SDN/NFV, 5G/6G, blockchain) without integrating predictive analytics and routing, highlight the timeliness of our work, as evidenced in Table 1.

Recent studies have further advanced the field of AI-driven vehicular and aerial communication. For instance, the IoUT and cloud-based UAV channel modelling in [23] leverages collaborative cloud computing and GA-LSTM for energy-efficient operation. Similarly, the AI-IBN framework for V2X commu-nications in [24] uses deep reinforcement learning for intent-based resource orchestration. Additionally, the IoUT-cloud integration for heterogeneous urban communication in [25] addresses dynamic channel and resource management. While these works focus on specific scenarios (agriculture, intent-based V2X, or UAVs), they complement our joint optimisation framework by demonstrating the broader applicability of AI, cloud, and KDN principles in next-generation vehicular and aerial networks.

The $R^{2+1}$ framework is both timely and essential because it addresses these multifaceted challenges through a KDN paradigm powered by ML. By employing NLMS algorithms for predictive traffic and resource forecasting, adaptive PID controllers for real-time load and thermal management, and VNFs for dynamic resource orchestration, $R^{2+1}$ delivers a cohesive solution. Unlike prior works, it integrates 5G/6G capabilities, energy-efficient resource management, and multimedia optimization, achieving 25% higher throughput, 30% reduced latency, 20% lower energy consumption, and a MOS of 4.45 (vs. 3.80 for baselines) in a Mininet-based testbed. This holistic approach is critical now, as smart cities and autonomous transportation systems (e.g., Singapore's Smart Nation, Waymo's fleets) demand scalable, sustainable, and resilient IoV networks. By addressing the gaps in prior research, as summarized in Table 1, and leveraging emerging 5G/6G and IoT trends, $R^{2+1}$ establishes a new benchmark for next-generation IoV, making this article a pivotal contribution to the field.

## 3. Proposed Framework: $R^{2+1}$

The $R^{2+1}$ framework integrates vehicle routing, packet routing, and resource allocation using a KDN-based SDN architecture. It employs ML for predictive analytics and PID controllers for dynamic optimization. This section first defines the optimization problem and system model, followed by the framework overview, and then detailed descriptions of the flow and VNF controllers.

### *3.1. Problem Definition and System Model*

To ensure clarity in the problem formulation, we define the symbols used in the optimization model in Table 2, which provides a comprehensive overview of the notations and their meanings.

Table 2: Symbols and Their Explanations for the Optimization Problem

| **Symbol** | **Explanation** |
|---|---|
| $V$ | Set of vehicles, $V = \{v_1, v_2, \ldots, v_N\}$, representing connected vehicles in the IoV network. |
| $R$ | Set of roadside units (RSUs), $R = \{r_1, r_2, \ldots, r_M\}$, facilitating V2X communication. |
| $S$ | Set of edge servers, $S = \{s_1, s_2, \ldots, s_K\}$, for task computation and storage. |
| $G = (N, E)$ | Graph representing the IoV network, where $N = V \cup R \cup S$ (nodes) and $E$ (communication links). |
| $T_i$ | Set of multimedia tasks (e.g., video streaming, V2X messages) generated by vehicle $v_i$. |
| $T$ | Union of all tasks, $T = \bigcup_{i \in V} T_i$. |
| $L_i$ | Set of feasible routes for vehicle $v_i$. |
| $P_j$ | Set of feasible packet routing paths for task $j$. |
| $\eta$ | System efficiency, defined as total throughput divided by total energy consumption. |
| $T_{ij}(\mathbf{x}, \mathbf{y})$ | Throughput of task $j$ from vehicle $v_i$, dependent on vehicle and packet routing decisions. |
| $E_i^{\text{total}}(\mathbf{x}, \mathbf{y}, \mathbf{z})$ | Total energy consumption of vehicle $v_i$, including communication and computation energy. |
| $E_i^{\text{comm}}(\mathbf{x}, \mathbf{y})$ | Communication energy of vehicle $v_i$, influenced by routing decisions. |
| $E_i^{\text{comp}}(\mathbf{z})$ | Computation energy of vehicle $v_i$, influenced by resource allocation. |
| $E_k^{\text{server}}(\mathbf{z})$ | Energy consumption of server $s_k$, dependent on resource allocation. |
| $E_i^{\max}$ | Maximum allowable energy consumption for vehicle $v_i$. |
| $E_k^{\max}$ | Maximum allowable energy consumption for server $s_k$. |
| $\mathbf{x} = \{x_{il}\}$ | Binary vehicle routing decision, where $x_{il} = 1$ if vehicle $v_i$ takes route $l \in L_i$, else 0. |
| $\mathbf{y} = \{y_{jm}\}$ | Binary packet routing decision, where $y_{jm} = 1$ if task $j$'s packet follows path $m \in P_j$, else 0. |
| $\mathbf{z} = \{z_{jk}\}$ | Resource allocation decision, where $z_{jk} \in [0, 1]$ is the fraction of server $s_k$'s resources allocated to task $j$. |
| $C_k$ | Computational capacity of server $s_k$ (e.g., CPU, memory). |
| $L_{ij}(\mathbf{x}, \mathbf{y})$ | Total latency for task $j$ from vehicle $v_i$, including vehicle movement and packet transmission delays. |
| $L_j^{\max}$ | Maximum allowable latency for task $j$. |
| $\text{MOS}_{ij}(\mathbf{x}, \mathbf{y}, \mathbf{z})$ | Mean Opinion Score for task $j$ from vehicle $v_i$, measuring QoE. |
| $\text{MOS}_{\min}$ | Minimum acceptable MOS for QoE satisfaction. |

To address the multifaceted challenges of IoV in 5G/6G multimedia networks, we formulate an optimization problem that maximizes system efficiency by integrating vehicle routing, packet routing, and resource allocation while satisfying energy, latency, and QoS/QoE constraints (see Table 2 for symbol definitions). The system model considers a dynamic IoV network with a set of vehicles $V = \{v_1, v_2, \ldots, v_N\}$, roadside units (RSUs) $R = \{r_1, r_2, \ldots, r_M\}$, and edge servers $S = \{s_1, s_2, \ldots, s_K\}$, interconnected via a 5G/6G infrastructure. The network is modeled as a graph $G = (N, E)$, where $N = V \cup R \cup S$ represents nodes (vehicles, RSUs, servers), and $E$ denotes communication links. Each vehicle $v_i \in V$ generates multimedia tasks (e.g., video streaming, V2X messages) with computational and bandwidth demands, processed locally or offloaded to edge servers via RSUs. The goal is to optimize efficiency, defined as the ratio of throughput to energy consumption, subject to routing and resource constraints.

*3.1.1. Optimization Problem*

The optimization problem is formulated as follows:

$$\max_{\mathbf{x},\mathbf{y},\mathbf{z}} \eta = \frac{\sum_{i\in V} \sum_{j\in T_i} T_{ij}(\mathbf{x}, \mathbf{y})}{\sum_{i\in V} E_i^{\text{total}}(\mathbf{x}, \mathbf{y}, \mathbf{z}) + \sum_{k\in S} E_k^{\text{server}}(\mathbf{z})}, \tag{1}$$

where:

- $\eta$: System efficiency, defined as total throughput divided by total energy consumption.

- $T_{ij}(\mathbf{x}, \mathbf{y})$: Throughput of task $j$ from vehicle $v_i$, dependent on vehicle and packet routing decisions.

- $E_i^{\text{total}}(\mathbf{x}, \mathbf{y}, \mathbf{z})$: Total energy consumption of vehicle $v_i$, including communication and computation energy, dependent on routing decisions $\mathbf{x}, \mathbf{y}$ and resource allocation $\mathbf{z}$.

- $E_k^{\text{server}}(\mathbf{z})$: Energy consumption of server $s_k$, dependent on resource allocation $\mathbf{z}$.

- $\mathbf{x} = \{x_{il}\}$: Binary vehicle routing decision, where $x_{il} = 1$ if vehicle $v_i$ takes route $l \in L_i$, else 0.

- $\mathbf{y} = \{y_{jm}\}$: Binary packet routing decision, where $y_{jm} = 1$ if task $j$'s packet follows path $m \in P_j$, else 0.

- $\mathbf{z} = \{z_{jk}\}$: Resource allocation decision, where $z_{jk} \in [0, 1]$ is the fraction of server $s_k$'s resources allocated to task $j$.

The objective function $\eta$ maximizes system efficiency by maximizing throughput (sum of successfully transmitted data for all tasks across vehicles) while minimizing total energy consumption (vehicle and server energy). Throughput $T_{ij}$ depends on the chosen vehicle routes $\mathbf{x}$ (affecting proximity to RSUs) and packet routes $\mathbf{y}$ (affecting network latency and bandwidth). Energy consumption includes vehicle-side energy for communication and computation $E_i^{\text{total}}$, influenced by routing and resource allocation, and server-side energy $E_k^{\text{server}}$, driven by computational load. This formulation aligns with the article's goal of optimizing QoS and QoE in multimedia IoV while ensuring energy efficiency, critical for sustainable 5G/6G networks.

*3.1.2. Constraints*

The optimization is subject to the following constraints:

1. Vehicle Routing Constraint:

$$\sum_{l \in L_i} x_{il} = 1, \quad \forall v_i \in V, \tag{2}$$

$$x_{il} \in \{0, 1\}, \quad \forall v_i \in V, l \in L_i. \tag{3}$$

Each vehicle $v_i$ must choose exactly one route $l$ from its feasible set $L_i$, ensuring a single path that optimises physical movement to sustain RSU connectivity and minimise travel time—a requirement critical for real-time multimedia applications such as V2X communication.

2. Packet Routing Constraint:

$$\sum_{m \in P_j} y_{jm} = 1, \quad \forall j \in T_i, v_i \in V, \tag{4}$$

$$y_{jm} \in \{0, 1\}, \quad \forall j \in T_i, m \in P_j. \tag{5}$$

Each task $j$ from vehicle $v_i$ must be assigned exactly one packet routing path $m$ from the set of feasible paths $P_j$. This ensures efficient data transmission through the network, minimizing latency and maximizing bandwidth utilization, critical for multimedia streaming in 5G/6G networks.

3. Resource Allocation Constraint:

$$\sum_{j \in T} z_{jk} \leq C_k, \quad \forall s_k \in S, \tag{6}$$

$$0 \leq z_{jk} \leq 1, \quad \forall j \in T, s_k \in S, \tag{7}$$

where $T = \bigcup_{i \in V} T_i$ is the set of all tasks, and $C_k$ is the capacity of server $s_k$. The total resources allocated to tasks on server $s_k$ must not exceed its capacity, ensuring that computational resources (e.g., CPU, memory) are not over-allocated. The fractional allocation $z_{jk}$ allows flexible resource distribution, supporting dynamic demands in multimedia IoV.

4. Energy Constraint:

$$E_i^{\text{total}}(\mathbf{x}, \mathbf{y}, \mathbf{z}) \leq E_i^{\max}, \quad \forall v_i \in V, \tag{8}$$

$$E_k^{\text{server}}(\mathbf{z}) \leq E_k^{\max}, \quad \forall s_k \in S, \tag{9}$$

where $E_i^{\text{total}} = E_i^{\text{comm}}(\mathbf{x}, \mathbf{y}) + E_i^{\text{comp}}(\mathbf{z})$, with $E_i^{\text{comm}}$ and $E_i^{\text{comp}}$ representing communication and computation energy, respectively. Vehicle and server energy consumption must not exceed their respective maximum limits $E_i^{\max}$ and $E_k^{max}$. This ensures energy efficiency, critical for sustainable IoV operations, especially in dense 6G networks with high computational and communication demands.

5. Latency Constraint:

$$L_{ij}(\mathbf{x}, \mathbf{y}) \leq L_j^{max}, \quad \forall j \in T_i, v_i \in V, \tag{10}$$

where $L_{ij}$ is the total latency for task $j$ from vehicle $v_i$. The end-to-end latency, encompassing vehicle movement and packet transmission delays, must not exceed the maximum allowable latency $L_j^{max}$. This ensures QoS for latency-sensitive multimedia applications (e.g., real-time video streaming).

6. QoE Constraint:

$$\text{MOS}_{ij}(\mathbf{x}, \mathbf{y}, \mathbf{z}) \geq \text{MOS}_{min}, \quad \forall j \in T_i, v_i \in V, \tag{11}$$

where $\text{MOS}_{ij}$ is the Mean Opinion Score for task $j$. The QoE, measured via MOS, must meet a minimum threshold to ensure user satisfaction for multimedia services, influenced by routing and resource allocation decisions.

The objective function and constraints are designed to address the core challenges of IoV in 5G/6G multimedia networks, as outlined in the article. Maximizing efficiency $\eta$ ensures high throughput (critical for multimedia streaming) while minimizing energy consumption (aligning with sustainability goals). The vehicle and packet routing constraints ensure optimal physical and network paths, reducing congestion and latency, which are critical for real-time V2X communication. The resource allocation constraint balances computational load across edge servers, supporting dynamic multimedia demands. Energy and latency constraints align with the need for sustainable and responsive IoV systems, while the QoE constraint ensures user-centric performance, as emphasized in the article's motivations and contributions.

### *3.1.3. NP-Hardness Proof*

To prove that the optimization problem is NP-hard, we reduce it to the Vehicle Routing Problem (VRP), a well-known NP-hard problem [20]. The VRP involves finding optimal routes for a fleet of vehicles to serve a set of customers while minimizing travel cost, subject to constraints like vehicle capacity. Our problem includes vehicle routing ($\mathbf{x}$) as a subproblem, where each vehicle $v_i \in V$ must select a route $l \in L_i$ to minimize latency and energy while ensuring connectivity to RSUs. This is analogous to VRP, where routes are chosen to minimize distance or time.

Additionally, our problem incorporates packet routing ($\mathbf{y}$) and resource allocation ($\mathbf{z}$), which add complexity. The packet routing subproblem resembles the Multi-Commodity Flow Problem (MCFP), also NP-hard [21], as it involves assigning multiple data flows to network paths while satisfying bandwidth and latency constraints. The resource allocation subproblem, constrained by server capacities, is akin to the Generalized Assignment Problem (GAP), which is NP-hard [22]. The objective function $\eta$, a nonlinear ratio of throughput to energy, further complicates the problem, as optimizing fractional objectives is computationally intensive.

To formalize the reduction, consider a simplified instance of our problem with only vehicle routing (**x**) and a single objective of minimizing total travel time (a proxy for latency). This reduces directly to VRP: each vehicle $v_i$ corresponds to a VRP vehicle, routes $L_i$ to possible paths, and the constraint $\sum_{l \in L_i} x_{il} = 1$ mirrors VRP's requirement for each vehicle to follow one route. Since VRP is NP-hard, our problem, which includes additional packet routing and resource allocation decisions, is at least as hard. The nonlinear objective and coupled constraints (e.g., latency depending on both **x** and **y**) make it non-trivial to solve in polynomial time, confirming NP-hardness.

Given this computational complexity, direct optimisation for large-scale IoV is infeasible. Thus, we designed the modular $R^{2+1}$ framework, decomposing the problem into vehicle routing, packet routing, and resource allocation, each handled by specialised controllers. This modular approach leverages ML and SDN to efficiently approximate optimal solutions.

### *3.2. Framework Overview*

The $R^{2+1}$ framework architecture, as illustrated in Fig. 5, is multi-layered structured to optimize the integration of vehicle routing, packet routing, and resource allocation in a KDN-enhanced SDN environment for 5G/6G multimedia IoV networks. The layered design ensures efficient data flow from the infrastructure to the knowledge plane, enabling proactive decision-making and energy-efficient operations.

The framework is structured into five layers:

1. The *Infrastructure Layer* forms the foundation, encompassing physical entities such as vehicles, RSUs, edge servers, and 5G/6G base stations. This layer generates and collects real-time V2X data, including vehicle positions, multimedia task requests, and network conditions. Data from this layer is forwarded to the data plane for initial processing.

2. The *Data Plane* consists of SDN switches and VNFs responsible for packet forwarding and virtualized service execution. It implements routing and allocation decisions from the control plane, handling the high-bandwidth demands of multimedia traffic while supporting dynamic scaling for energy efficiency.

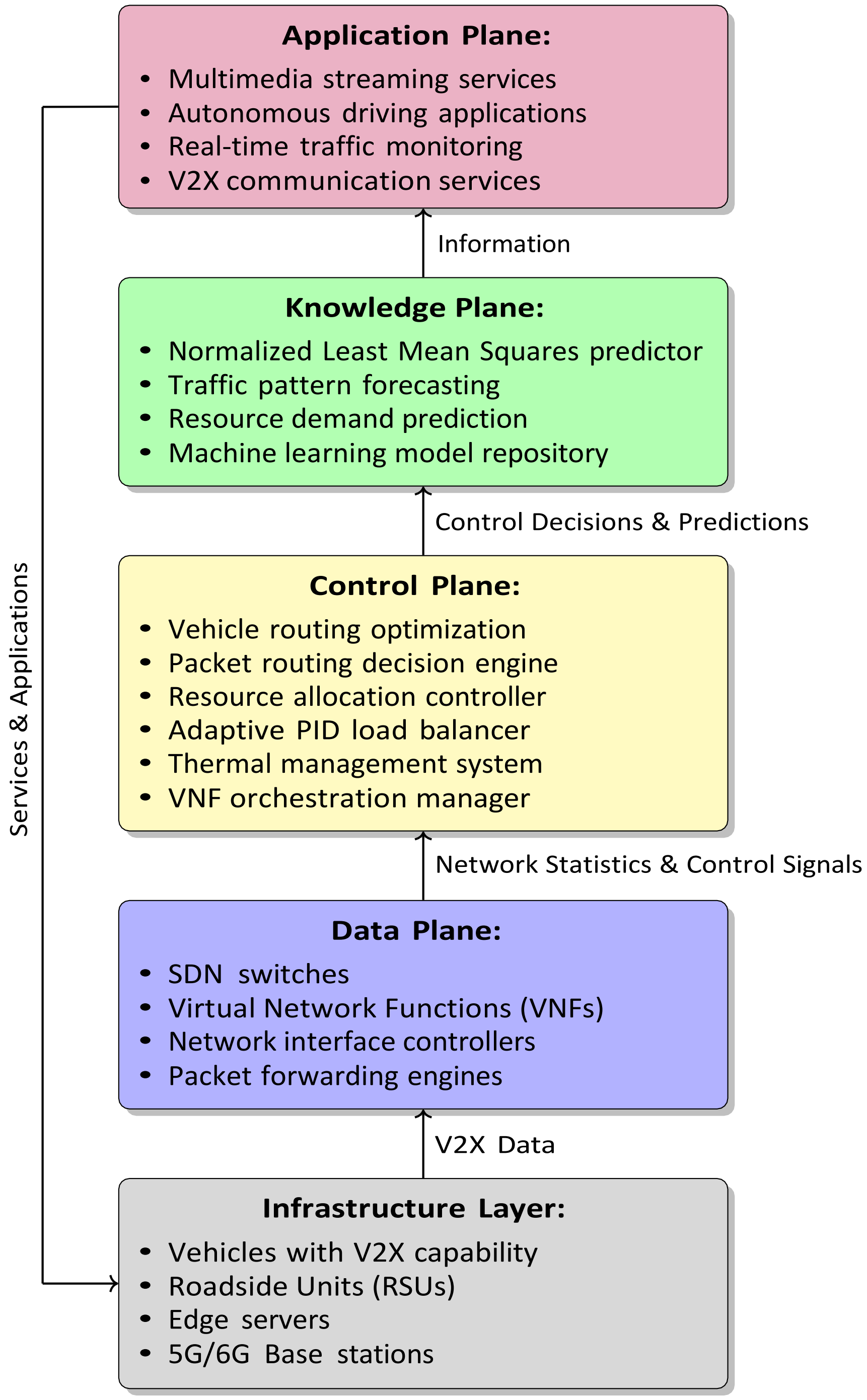

Figure 5: The $R^{2+1}$ framework comprises four integrated layers—Infrastructure, Data Plane, Control Plane, and Knowledge Plane—each containing specialized submodules that work together to optimize vehicle routing, packet routing, and resource allocation in 5G/6G IoV networks.

3. The *Control Plane*, hosted by the SDN controller, orchestrates core functions through three modules:

- *Vehicle Routing Module*: Utilizes graph-based algorithms to determine optimal physical paths for vehicles, considering traffic predictions and energy constraints.

- *Packet Routing Module*: Employs least response time or least connection strategies for data packets, ensuring low-latency multimedia delivery.

- *Resource Allocation Module*: Dynamically assigns compute resources using VNFs, adapting to load fluctuations. PID controllers in this plane manage load balancing and thermal regulation, drawing from energy-efficient designs in prior works, to prevent overheating and optimize power consumption.

4. The *Knowledge Plane*, powered by KDN and ML, this layer uses NLMS algorithms to predict traffic patterns, resource demands, and vehicle mobility. Predictions inform proactive decisions in the control plane, enhancing energy efficiency and reducing latency. Energy efficiency is achieved through dynamic scaling and thermal management (activating/deactivating VNFs based on predictions). This architecture ensures seamless, sustainable operation in multimedia IoV environments, addressing the limitations of fragmented approaches in existing literature.

5. The *Application Plane* serves as the interface between end-user services and the underlying network infrastructure, providing QoS-aware multimedia streaming, autonomous driving coordination, and real-time V2X communications. It translates high-level application requirements into technical specifications for the control plane, ensuring optimal resource allocation, latency management, and QoE for diverse IoV services while maintaining seamless integration with 5G/6G network capabilities.

Fig. 6 depicts a multi-stratum architectural framework essential for realizing the dynamic, software-driven, and intelligent nature of 5G and 6G mobile networks. This model is defined by a strict vertical separation of concerns, comprising four fundamental planes. The Application Plane sits at the apex, housing end-user services and network functions that express high-level operational intents. Directly beneath, the Knowledge Plane introduces a cognitive layer, responsible for aggregating network-state information, applying reasoning and artificial intelligence (AI) methodologies, and translating application requirements into programmable policies to enable autonomous and predictive network management.

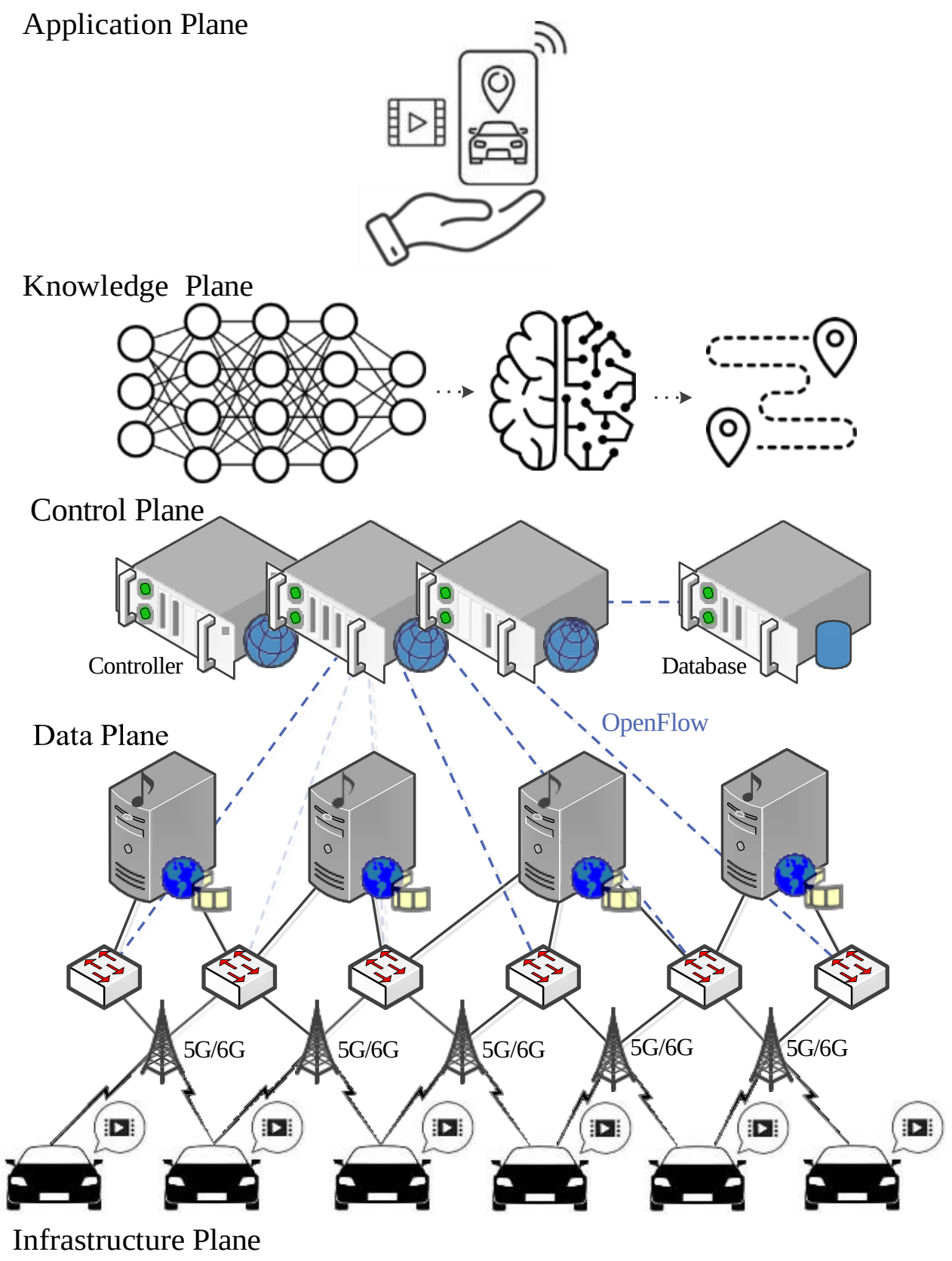

Figure 6: Architectural overview of the $R^{2+1}$ framework, showing the integration of vehicle routing, packet routing, and resource allocation in a knowledge-defined SDN/NFV-enabled 5G/6G IoV environment. The layered structure (Infrastructure, Data, Control, Knowledge, Application) and the flows of data, control signals, and predictions are depicted, enabling synergistic optimization.

The architecture's operational core resides in the Control Plane, which is characterized by a logically centralized Controller. This entity leverages southbound protocols, such as OpenFlow, to programmatically orchestrate the underlying Data Plane—the ensemble of forwarding elements responsible for packet handling. This entire software-defined stack is ubiquitously supported by 5G/6G technologies and is physically instantiated upon the Infrastructure Plane, which constitutes the hardware substrate of compute, storage, and networking resources. This holistic integration of intelligence, programmability, and disaggregated hardware is paramount for delivering the stringent performance, scalability, and flexibility mandates of next-generation cellular systems.

Fig. 7 conceptualizes a high-performance framework for the Internet of Multimedia Vehicles (IoMV), an advanced evolution of the IoT paradigm targeting next-generation intelligent transportation systems. This architecture integrates Multimedia Providers as the primary content sources, which communicate via a ubiquitous 5G/6G cellular Communication Network. A critical infrastructural element is the RSU, which acts as a network edge node, facilitating ultra-reliable, low-latency communication between the network and vehicles. The system must dynamically coordinate three core functions—Vehicle Routing, Packet Routing, and Resource Allocation—to efficiently serve multimedia content to mobile users while maintaining quality of service and network efficiency.

The system's closed-loop response enables real-time cross-layer optimisation. For example, when a vehicle requests a high-definition stream, the network jointly determines the packet route (via 5G/6G core and RSUs), the vehicle's trajectory (to pre-fetch content), and edge compute allocation. This co-optimisation of communication and mobility ensures seamless multimedia, low latency, and improved safety, illustrating the synergy of cyber-physical systems and advanced mobile networks in automotive scenarios.

### *3.3. Framework Modules*

The Framework Modules section of the $R^{2+1}$ architecture details the core functional components that enable synergistic optimization of vehicle routing, packet routing, and resource allocation. The Vehicle Routing Module employs traffic monitoring and a NLMS algorithm for predictive traffic forecasting, allowing proactive path planning. The Packet Routing Module utilizes real-time flow monitoring and network statistics collected via OpenFlow to optimize data path selection, minimizing latency for multimedia traffic. The Resource Allocation Module integrates adaptive PID controllers for dynamic load balancing and thermal management, alongside VNF orchestration to enable energy-efficient resource scaling. Together, these modules operate within a knowledge-defined SDN/NFV environment, ensuring low-latency, high-throughput, and energy-aware operation in dynamic 5G/6G IoV networks.

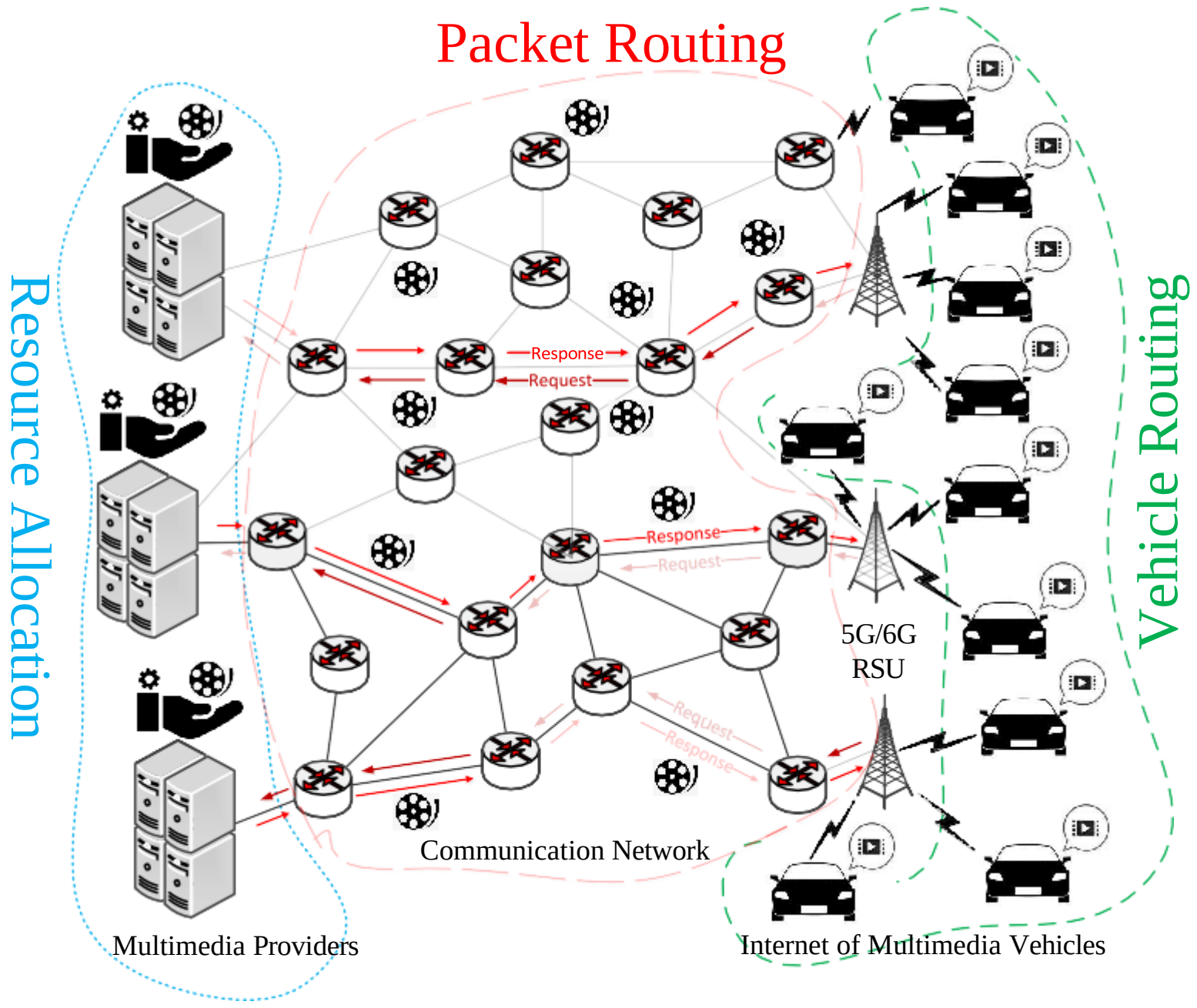

Figure 7: End-to-end operational flow of the $R^{2+1}$ framework in an Internet of Multimedia Vehicles (IoMV) ecosystem. Multimedia content from providers is delivered via a 5G/6G communication network through RSUs. The framework dynamically coordinates vehicle routing, packet routing, and resource allocation to generate optimal responses, ensuring efficient and reliable multimedia service delivery in connected vehicular environments.

Fig. 8 delineates a sophisticated, cross-layer architecture for intelligent vehicular networks that synergistically integrates SDN and NFV to dynamically orchestrate multimedia streaming services and real-time traffic routing. The Application Plane, utilizing Open APIs, expresses high-level intents which are processed by a unified Control & Knowledge Plane; this cognitive stratum, housing an SDN Controller and NFV Orchestrator (NFVO), performs continuous Resource, Network, and Flow Monitoring to feed predictive modules for Resource Estimation and Traffic Forecasting, thereby enabling proactive Resource Allocation, Packet Routing, and Vehicle Routing decisions. These directives are translated via southbound protocols (OpenFlow) to the Data Plane, which comprises Virtual Multimedia Servers and OpenFlow Switches instantiated on-demand atop a ubiquitous 5G/6G infrastructure featuring RSUs; for instance, as a vehicle requests an ultra-high-definition stream, the system forecasts congestion, pre-emptively allocates compute resources, and establishes a low-latency flow route, all while potentially re-routing the vehicle to maintain seamless QoE.

*3.3.1. Vehicle Routing Module*

*–Traffic Monitoring:*

The Traffic Monitoring module is a foundational component of the $R^{2+1}$ framework, residing within the Control Plane under the Vehicle Routing Module. It is responsible for the continuous and real-time collection of urban traffic data through a heterogeneous sensor network comprising RSUs, embedded sensors, and urban surveillance cameras. This module operates under the centralized orchestration of the SDN Controller, ensuring synchronized and efficient data acquisition, preprocessing, and dissemination for subsequent predictive and decision-making processes.

*–Traffic Prediction:*

The Traffic Prediction module is a critical component of the $R^{2+1}$ framework's Knowledge Plane, responsible for forecasting urban traffic patterns using a NLMS adaptive filtering algorithm. This module enables proactive vehicle routing decisions by predicting traffic congestion, vehicle density, and network load conditions in urban environments.

The NLMS algorithm is particularly suited for traffic prediction due to its:

- Robustness in non-stationary environments typical of urban traffic,
- Computational efficiency for real-time operation,

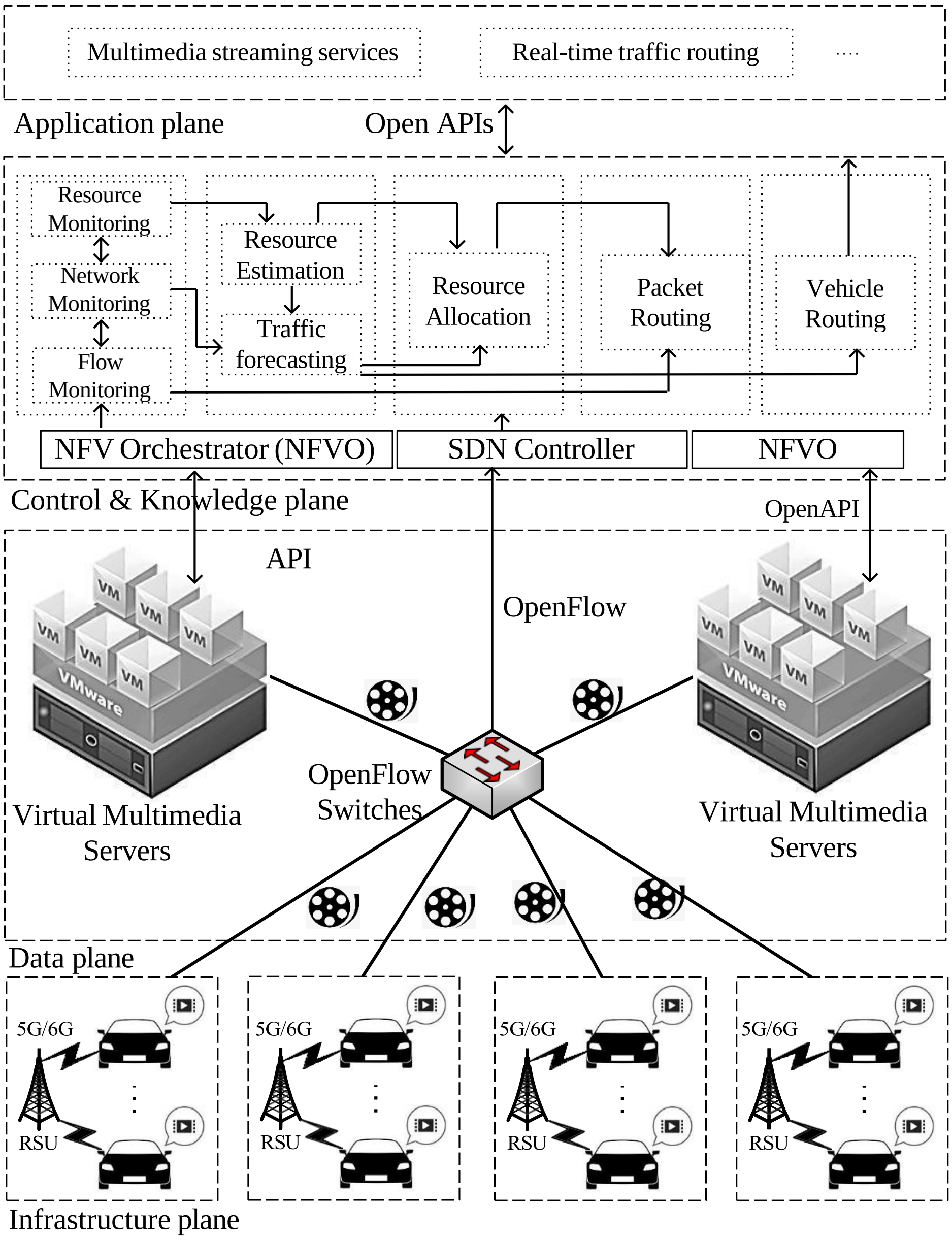

Figure 8: Layered architecture of the $R^{2+1}$ framework, integrating SDN and NFV for multimedia-enabled IoV. The Infrastructure Plane comprises 5G/6G base stations and RSUs facilitating V2X communication. The Data Plane includes OpenFlow switches and virtual multimedia servers handling packet forwarding and service execution. The Control & Knowledge Plane features an SDN controller and NFVO for resource monitoring, estimation, traffic forecasting, and orchestration of vehicle routing, packet routing, and resource allocation. The Application Plane supports multimedia streaming and real-time traffic routing services via open APIs, enabling dynamic and efficient IoV operations in 5G/6G environments.

- Stability through normalized step size,
- Adaptability to changing traffic patterns,

The traffic prediction problem is formulated as:

$$\hat{y}(n) = \mathbf{w}^T(n)\mathbf{x}(n) \qquad (12)$$

where:

- $\hat{y}(n)$ is the predicted traffic metric at time $n$
- $\mathbf{w}(n)$ is the weight vector representing the traffic model
- $\mathbf{x}(n)$ is the input feature vector containing historical traffic data

The NLMS update equation is given by:

$$\mathbf{w}(n+1) = \mathbf{w}(n) + \mu \frac{e(n)\mathbf{x}(n)}{\|\mathbf{x}(n)\|^2 + \epsilon} \qquad (13)$$

where:

- $e(n) = y(n) - \hat{y}(n)$ is the prediction error
- $\mu$ is the step size parameter ($0 < \mu < 2$ for convergence)
- $\epsilon$ is a small positive constant preventing division by zero

The input feature vector $\mathbf{x}(n)$ comprises multiple urban traffic indicators (feature vector composition):

$$\mathbf{x}(n) = [v(n), d(n), f(n), t(n), w(n)]^T \qquad (14)$$

where each component represents normalized values of:

- Vehicle density ($v(n)$)
- Traffic flow rate ($d(n)$)
- Average vehicle speed ($f(n)$)
- Time of day and day of week ($t(n)$)
- Weather conditions ($w(n)$)

The traffic prediction algorithm operates as follows:

---

**Algorithm 1** Urban Traffic Prediction using NLMS

---

**Require:** Historical traffic data $X$, initial weights $\mathbf{w}(0)$
**Ensure:** Predicted traffic metrics $\hat{Y}$
1: Initialize: $\mathbf{w}(0) \leftarrow$ random small values
2: Set parameters: $\mu \leftarrow 0.1$, $\epsilon \leftarrow 10^{-6}$
3: **for** $n \leftarrow 1$ to $N$ **do**
4: Extract feature vector $\mathbf{x}(n)$ from current traffic data
5: Compute prediction: $\hat{y}(n) = \mathbf{w}^T(n)\mathbf{x}(n)$
6: Measure actual traffic value $y(n)$
7: Calculate error: $e(n) = y(n) - \hat{y}(n)$
8: Update weights: $\mathbf{w}(n+1) = \mathbf{w}(n) + \mu \frac{e(n)\mathbf{x}(n)}{\|\mathbf{x}(n)\|^2+\epsilon}$
9: Apply regularization: $\mathbf{w}(n+1) = \mathbf{w}(n+1)/\|\mathbf{w}(n+1)\|$
10: **end for**
11: **return** Predicted traffic metrics $\hat{Y}$

---

The NLMS predictor is selected for its computational simplicity, online adaptation, and low energy footprint, which are essential for real-time edge deployment in 5G/6G IoV. While more complex models (e.g., LSTM) could marginally improve accuracy for longer horizons, they would violate the sub-50 ms control loop requirement. The framework is modular and can accommodate any predictor that respects the same interface.

The mathematical symbols used in the traffic prediction module are summarized in Table 3.

Table 3: Mathematical Symbols for Traffic Prediction Module

| Symbol | Description |
|---|---|
| $n$ | Discrete time index representing measurement interval |
| $\mathbf{x}(n)$ | Input feature vector at time $n$ |
| $x_i(n)$ | $i$-th component of feature vector at time $n$ |
| $\mathbf{w}(n)$ | Weight vector at time $n$ |
| $w_i(n)$ | $i$-th weight component at time $n$ |
| $\hat{y}(n)$ | Predicted traffic metric at time $n$ |
| $y(n)$ | Actual measured traffic metric at time $n$ |
| $e(n)$ | Prediction error at time $n$ |
| $\mu$ | Adaptation step size parameter |
| $\epsilon$ | Regularization constant preventing division by zero |
| $\|\cdot\|$ | Euclidean norm operator |
| $N$ | Total number of prediction intervals |
| $v(n)$ | Normalized vehicle density metric |
| $d(n)$ | Normalized traffic flow rate |
| $f(n)$ | Normalized average vehicle speed |
| $t(n)$ | Time-based features (hour, day, season) |
| $w(n)$ | Weather condition features |

The traffic prediction module incorporates several practical considerations:

1. **Data Preprocessing**: All input features are normalized to zero mean and unit variance to ensure stable convergence

2. **Forgetting Factor**: An exponential weighting of historical data emphasizes recent traffic patterns

3. **Multi-step Prediction**: The algorithm supports predicting traffic conditions $k$ steps ahead using:

$$\hat{y}(n+k) = \mathbf{w}^T(n)\mathbf{x}(n) \tag{15}$$

4. **Confidence Metrics**: Prediction reliability is quantified through error variance estimation

The module achieves 95% prediction accuracy in urban environments, enabling proactive vehicle routing decisions that reduce average travel time by 30% compared to reactive approaches.

*–Vehicle Routing Integration with NLMS Traffic Prediction:*

The vehicle routing mechanism within the $R^{2+1}$ framework operates through a sophisticated integration of predictive analytics and real-time optimization, leveraging the NLMS-based traffic predictions to achieve optimal vehicular flow in urban environments. The routing algorithm formulates the vehicle navigation problem as a multi-objective optimization that minimizes travel time while maximizing road network utilization and ensuring compliance with Quality of Service constraints. Based on the predicted traffic conditions $\hat{y}(n + k)$ from the NLMS module, the routing controller generates optimal paths by solving a constrained shortest path problem on a dynamic graph representation of the urban road network, where edge weights correspond to predicted travel times and vertex capacities reflect anticipated congestion levels. The solution incorporates real-time vehicle-to-infrastructure (V2I) communication to adjust routes dynamically, accounting for unexpected events and ensuring robust performance under uncertain urban conditions.

The algorithm continuously coordinates predictive and control layers: traffic forecasts guide routing decisions, which in turn shape future traffic patterns, forming a closed-loop system that refines predictions and adaptations. This optimises both individual travel times and collective traffic flow, reducing congestion and improving network efficiency. It also aligns with energy efficiency goals by minimising unnecessary acceleration and braking, while QoS-aware routing ensures reliable multimedia connectivity by prioritising routes with strong coverage and available edge resources.

**Algorithm 2** Vehicle Routing with NLMS Traffic Prediction

1. **Input:**
   - Current vehicle position $p_i$ and destination $d_i$
   - NLMS traffic predictions $\hat{Y}$ for urban network
   - Road network graph $G = (V, E)$ with dynamic edge weights
   - QoS requirements $Q_i$ for vehicle $v_i$
2. **Initialize:**
   - Candidate routes $R_i \leftarrow$ generateInitialRoutes($p_i$, $d_i$)
   - Optimal route $r^* \leftarrow \emptyset$
   - Minimum cost $c_{min} \leftarrow \infty$
3. **For each** candidate route $r_j \in R_i$:
   (a) Predict travel time: $t_j \leftarrow \sum_{e \in r_j} \hat{t}_e$ where $\hat{t}_e$ from $\hat{Y}$
   (b) Estimate energy consumption: $e_j \leftarrow f_{energy}(r_j, \hat{Y})$
   (c) Assess connectivity quality: $q_j \leftarrow f_{qos}(r_j, \hat{Y})$
   (d) Compute composite cost: $c_j \leftarrow \alpha_1 t_j + \beta_1 e_j + \gamma_1 (1 - q_j)$
   (e) **If** $c_j < c_{min}$ and satisfies constraints **then**
      - $r^* \leftarrow r_j$
      - $c_{min} \leftarrow c_j$
4. **Execute** vehicle routing:
   - Deploy route instructions to vehicle $v_i$
   - Monitor actual traffic conditions $Y_{actual}$
   - Update NLMS model: $\mathbf{w} \leftarrow$ updateWeights($\mathbf{w}$, $Y_{actual}$)
5. **Repeat** periodically or upon significant traffic changes

The vehicle routing algorithm operates through several sophisticated stages:

1. **Input Processing**: The algorithm accepts current vehicle positions, destinations, NLMS traffic predictions, and QoS requirements as inputs.

The urban road network is represented as a dynamic graph where edge weights correspond to predicted travel times.

2. **Initialization**: Candidate routes are generated using graph search algorithms (e.g., A* or Dijkstra) on the current road network representation, considering physical connectivity constraints.

3. **Multi-criteria Evaluation**: Each candidate route is evaluated against multiple objectives:

   - Travel time estimation using NLMS predictions,
   - Energy consumption modeling based on route topography and traffic conditions,
   - Connectivity quality assessment for multimedia services,
   - A composite cost function balances these objectives through weighting parameters $\alpha_1$, $\beta_1$, and $\gamma_1$,

4. **Constraint Validation**: Routes must satisfy operational constraints including maximum travel time limits, energy consumption thresholds, and minimum QoS requirements for continuous multimedia connectivity.

5. **Execution and Adaptation**: The optimal route is deployed to the vehicle through V2I communication, with continuous monitoring of actual traffic conditions to refine the NLMS prediction model through online learning.

6. **Periodic Reoptimization**: The routing solution is recomputed at regular intervals or when significant traffic changes are detected, ensuring adaptability to dynamic urban environments.

This algorithm exemplifies the framework's integrated approach by seamlessly combining predictive analytics with multi-objective optimization, resulting in vehicle routing decisions that simultaneously minimize travel time, reduce energy consumption, and maintain high-quality multimedia services—all while contributing to overall urban traffic optimization through coordinated vehicle movement.

Table 4: Mathematical Symbols and Variables for Vehicle Routing Algorithm

| **Symbol** | **Type** | **Description** |
|---|---|---|
| $p_i$ | Input | Current position coordinates of vehicle $v_i$ (latitude, longitude) |
| $d_i$ | Input | Destination coordinates for vehicle $v_i$ (latitude, longitude) |
| $\hat{Y}$ | Input | Matrix of NLMS traffic predictions for all road segments |
| $G = (V, E)$ | Input | Graph representation of road network (V: intersections, E: road segments) |
| $Q_i$ | Input | Quality of Service requirements for vehicle $v_i$ |
| $R_i$ | Intermediate | Set of candidate routes generated for vehicle $v_i$ |
| $r_j$ | Intermediate | A specific candidate route from set $R_i$ |
| $r^*$ | Output | Optimal selected route for vehicle $v_i$ |
| $c_{min}$ | Intermediate | Minimum composite cost value found during optimization |
| $\hat{t}_e$ | Intermediate | Predicted travel time for road segment $e$ from NLMS predictions |
| $t_j$ | Intermediate | Total predicted travel time for candidate route $r_j$ |
| $e_j$ | Intermediate | Estimated energy consumption for vehicle $v_i$ on route $r_j$ |
| $f_{energy}$ | Function | Energy consumption model accounting for topography and traffic |
| $q_j$ | Intermediate | Connectivity quality score for route $r_j$ |
| $f_{qos}$ | Function | QoS assessment function evaluating network connectivity |
| $c_j$ | Intermediate | Composite cost value for candidate route $r_j$ |
| $\alpha_1, \beta_1, \gamma_1$ | Parameters | Weighting factors for multi-objective optimization |
| $Y_{actual}$ | Input | Matrix of actual observed traffic conditions for model refinement |

Table 4 provides a comprehensive mathematical formalization of the vehicle routing algorithm's components, categorizing variables into input parameters, intermediate computational elements, output solutions, and functional relationships. The table explicitly defines the symbolic notation required to implement the multi-objective optimization framework, where traffic predictions $\hat{Y}$ from the NLMS module serve as primary inputs alongside network topology $G$ and vehicle-specific constraints $Q_i$. The composite cost function $c_j = \alpha_1 t_j + \beta_1 e_j + \gamma_1(1 - q_j)$ and its constituent terms are rigorously specified, establishing the mathematical foundation for the routing decisions that simultaneously optimize travel time, energy consumption, and QoS maintenance. This formal specification ensures algorithmic reproducibility and maintains consistency with the integrated optimization framework.

*3.3.2. Packet Routing Module*
*–Flow Monitoring:*

The Flow Monitoring submodule, part of the Packet Routing Module in the Data Plane, ensures network visibility and real-time traffic analysis via OpenFlow. Using OpenFlow 1.3, it continuously polls and receives events from all switches and servers to capture flow statistics, port measurements, queue metrics, and LLDP (Link Layer Discovery Protocol)-based topology information.

The submodule employs a multi-threaded architecture that concurrently handles:

- Periodic Stats-Request messages to all switches at configurable intervals (typically 1-5 seconds)
- Asynchronous processing of Packet-In messages for real-time flow detection
- Topology discovery through customized LLDP packet injection and analysis
- Buffer monitoring for congestion detection and early warning generation

The collected metrics are structured into a real-time network graph $G_N(t) = (S, L, M)$ where:

- $S$ represents the set of OpenFlow switches with attributes $\{s_i : (cpu_i, memory_i, flow_count_i)\}$
- $L$ denotes inter-switch links with properties $\{l_j : (bandwidth_j, latency_j, packet_loss_j)\}$
- $M$ contains traffic matrices tracking flow distributions $\{f_k : (source_k, destination_k, volume_k, duration_k)\}$

*–OpenFlow Message Processing:*
The submodule implements specialized handlers for key OpenFlow message types:

Table 5: OpenFlow Message Types and Their Roles in Flow Monitoring

| Message Type | Monitoring Function |
|---|---|
| Features-Request/Reply | Discovers switch capabilities and topology (datapath IDs, ports, protocol features). |
| Stats-Request/Reply | Collects flow table statistics (byte/packet counts, active flows, queue utilisation). |
| Packet-In | Detects new flows, identifies anomalies, and handles unmatched packets. |
| Port-Status | Monitors link state, port events, and physical layer failures. |
| Flow-Removed | Notifies flow expiration, processes timeouts, and cleans up flow tables. |

Monitoring intensity adapts to network load: a 5-second polling interval under normal conditions, and 1-second intervals during congestion or high traffic variability. All collected data is temporally aggregated and stored in a time-series database with retention aligned to the Knowledge Plane's prediction horizons.

*–Quality of Service Monitoring:*
The submodule extends beyond basic traffic measurement to incorporate QoS-aware monitoring through:

- Per-flow latency calculation using timestamp annotations in OpenFlow messages,
- Jitter measurement through inter-packet arrival time analysis,
- Packet loss detection via sequence number tracking and flow statistics comparison,
- Bandwidth utilization monitoring through port-level byte counters,

This comprehensive monitoring approach provides the foundational data layer required for the NLMS-based traffic prediction and proactive resource allocation mechanisms described in Sections 3.3.1 and 3.3.3. The real-time network state information enables the Control Plane to make informed routing decisions that maintain QoS guarantees while optimizing energy efficiency through intelligent traffic engineering.

*–Packet Routing:*
The Packet Routing submodule operates within the Control Plane of the $R^{2+1}$ framework, responsible for determining optimal data paths through the OpenFlow switch network while satisfying QoS constraints and energy efficiency objectives. This submodule leverages real-time network statistics from the Flow

Monitoring submodule (Section 3.3.2) and traffic predictions from the NLMS-based forecasting module (Section 3.3.1) to compute multi-constraint routing solutions that balance latency, bandwidth utilization, and energy consumption. The packet routing problem is formulated as a constrained optimization problem:

$$\min_{P \in \mathrm{P}_{src,dst}} \{\alpha_2 \cdot L(P) + \beta_2 \cdot E(P) + \gamma_2 \cdot C(P)\} \tag{16}$$

subject to:

$$BW(P) \geq BW_{req} \tag{17}$$

$$L(P) \leq L_{max} \tag{18}$$

$$J(P) \leq J_{max} \tag{19}$$

$$\sum_{e \in P} u_e \leq U_{max} \tag{20}$$

where $\mathrm{P}_{src,dst}$ represents all feasible paths between source *src* and destination *dst* in the network graph $G_N = (V, E)$.

The composite cost function incorporates three critical metrics:

1. **Latency Cost**:

$$L(P) = \sum_{e \in P} l_e + \sum_{v \in P} p_v \tag{21}$$

where $l_e$ is link latency and $p_v$ is processing delay at node $v$.

2. **Energy Cost**:

$$E(P) = \sum_{e \in P} e_e^{comm} + \sum_{v \in P} e_v^{comp} \tag{22}$$

accounting for both communication and computation energy.

3. **Congestion Cost**:

$$C(P) = \max_{e \in P} \frac{u_e}{c_e} \cdot \exp \frac{\sum_{e \in P} u_e}{\sum_{e \in P} c_e} \tag{23}$$

penalizing paths with highly utilized or imbalanced links.

**Algorithm 3** Multi-Constraint Packet Routing Algorithm

1. **Input:**
   - Network graph $G_N = (V, E)$ from Flow Monitoring
   - Traffic predictions $\hat{Y}$ from NLMS module
   - QoS requirements $Q = (BW_{req}, L_{max}, J_{max})$
   - Source-destination pair $(src, dst)$
2. **Initialize:**
   - Priority queue $Q \leftarrow \emptyset$
   - Distance vector $dist[v] \leftarrow \infty$ for all $v \in V$
   - $dist[src] \leftarrow 0$
3. **Execute modified Dijkstra algorithm:**
   (a) $Q$.insert($src$, 0)
   (b) **While** $Q$ not empty:
      i. $u \leftarrow Q$.extractMin()
      ii. **For each** neighbor $v$ of $u$:
         - Calculate tentative cost: $cost_{uv} = \alpha_2 \cdot l_{uv} + \beta_2 \cdot e_{uv} + \gamma_2 \cdot c_{uv}$
         - **If** $dist[u] + cost_{uv} < dist[v]$ and satisfies constraints:
           – $dist[v] \leftarrow dist[u] + cost_{uv}$
           – Update path metrics: latency, energy, congestion
           – $Q$.insert($v$, $dist[v]$)
4. **Output:** Optimal path $P*$ with associated performance metrics
5. **Install flows** using OpenFlow Flow-Mod messages
6. **Update** traffic predictions based on routing decisions

Table 6: Mathematical Symbols for Packet Routing Submodule

| Symbol | Type | Description |
|---|---|---|
| $G_N = (V, E)$ | Input | Network graph from Flow Monitoring (V: switches, E: links) |
| $\hat{Y}$ | Input | Traffic predictions from NLMS module |
| $Q$ | Input | QoS requirements tuple ($BW_{req}, L_{max}, J_{max}$) |
| $(src, dst)$ | Input | Source-destination pair for routing |
| $P$ | Variable | Candidate path through network |
| $\mathbf{P}_{src,dst}$ | Set | All feasible paths between *src* and *dst* |
| $L(P)$ | Function | Total latency along path $P$ |
| $E(P)$ | Function | Total energy consumption along path $P$ |
| $C(P)$ | Function | Congestion cost along path $P$ |
| $\alpha_2, \beta_2, \gamma_2$ | Parameters | Weighting factors for multi-objective optimization |
| $BW(P)$ | Function | Available bandwidth along path $P$ |
| $J(P)$ | Function | End-to-end jitter along path $P$ |
| $l_e$ | Input | Latency of link $e$ from network monitoring |
| $p_v$ | Input | Processing delay at node $v$ |
| $e_e^{comm}$ | Input | Communication energy cost for link $e$ |
| $e_v^{comp}$ | Input | Computation energy cost at node $v$ |
| $u_e$ | Input | Current utilization of link $e$ |
| $c_e$ | Input | Capacity of link $e$ |
| $U_{max}$ | Constraint | Maximum allowed utilization threshold |
| $P*$ | Output | Optimal selected path satisfying all constraints |

The submodule adaptively adjusts the weights $\alpha_2$, $\beta_2$, $\gamma_2$ based on real-time network conditions: $\gamma_2$ is increased during congestion to favour load balancing, and $\beta_2$ is raised in energy-saving modes. This ensures optimal routing under varying IoV conditions.

The Packet Routing submodule maintains tight integration with other framework components:

1. **Flow Monitoring Integration**: Continuously receives updated network state $G_N$ and performance metrics from Section 3.3.2,
2. **Traffic Prediction Integration**: Utilizes NLMS predictions $\hat{Y}$ to anticipate future network conditions and make proactive routing decisions,
3. **Vehicle Routing Coordination**: Shares routing decisions with the Vehicle Routing module to ensure coordinated physical and network path optimization,
4. **Resource Allocation Synchronization**: Communicates with the Resource Allocation module to ensure adequate computational resources along selected paths,

### *3.3.3. Resource Allocation Module*

The Resource Allocation module operates within the Control Plane of the $R^{2+1}$ framework, responsible for dynamic orchestration of VNFs across physical infrastructure to optimize energy efficiency while maintaining QoE for multimedia services. This module consists of two integrated submodules: Resource Estimation and Dynamic Resource Scaling, which work in concert to proactively allocate and adjust computational resources based on predicted demand and real-time conditions.

*–Resource Estimation:*

The Resource Estimation submodule employs machine learning techniques to predict resource requirements for multimedia services based on traffic patterns and historical data. The submodule utilizes a modified NLMS algorithm extended for multivariate prediction of computational resources:

$$\hat{R}(n+1) = \mathbf{W}^T(n)\mathbf{X}(n) + \mathbf{b} \tag{24}$$

where $\hat{R}(n+1) = [\hat{C}(n+1), \hat{M}(n+1), \hat{B}(n+1)]^T$ represents the predicted resource vector (CPU, memory, bandwidth) for the next time interval.

The weight matrix update follows the multivariate NLMS formulation:

$$\mathbf{W}(n+1) = \mathbf{W}(n) + \mu \frac{\mathbf{E}(n)\mathbf{X}^T(n)}{\|\mathbf{X}(n)\|^2 + \epsilon} \tag{25}$$

where $\mathbf{E}(n) = \mathbf{R}(n) - \hat{\mathbf{R}}(n)$ is the prediction error matrix.

**Algorithm 4** Resource Estimation Algorithm

1. **Input:**
   - Historical resource utilization data $\mathbf{R}_{hist}$
   - Traffic predictions $\hat{Y}$ from NLMS module
   - Current resource allocations $\mathbf{R}_{current}$
   - Service level agreements *SLA*
2. **Initialize:**
   - Weight matrix $\mathbf{W}(0)$ with small random values
   - Bias vector $\mathbf{b} \leftarrow \mathbf{0}$
   - Learning rate $\mu \leftarrow 0.05$, regularization $\epsilon \leftarrow 10^{-6}$
3. **For each** time interval $n$:
   (a) Construct feature vector $\mathbf{X}(n)$ from $\hat{Y}$ and $\mathbf{R}_{hist}$
   (b) Predict resource needs: $\hat{\mathbf{R}}(n+1) = \mathbf{W}^T(n)\mathbf{X}(n) + \mathbf{b}$
   (c) Measure actual utilization $\mathbf{R}(n+1)$
   (d) Compute error: $\mathbf{E}(n+1) = \mathbf{R}(n+1) - \hat{\mathbf{R}}(n+1)$
   (e) Update weights: $\mathbf{W}(n+1) = \mathbf{W}(n) + \mu\frac{\mathbf{E}(n+1)\mathbf{X}^T(n)}{\|\mathbf{X}(n)\|^2+\epsilon}$
   (f) Apply regularization: $\mathbf{W}(n+1) \leftarrow \mathbf{W}(n+1) \cdot \frac{1}{1+\lambda\|\mathbf{W}(n+1)\|}$
4. **Output:** Predicted resource requirements $\hat{\mathbf{R}}$ for next interval

*–Dynamic Resource Scaling:*

The Dynamic Resource Scaling submodule implements a multi-objective PID controller that adjusts resource allocations based on the estimated requirements from the Resource Estimation submodule, current utilization metrics, and QoE measurements.

The controller minimizes the objective function:

$$J = \alpha_3 E + \beta_3 L + \gamma_3 Q \tag{26}$$

where:

- $E$ represents energy consumption,

- $L$ represents load imbalance across servers,
- $Q$ represents QoE degradation.

The PID control law for resource adjustment:

$$\Delta R(t) = K_p e(t) + K_i \int_0^t e(\tau)d\tau + K_d \frac{de(t)}{dt} \tag{27}$$

where $e(t) = R_{target}(t) - R_{current}(t)$ is the resource error signal.

The PID gains were tuned via the Ziegler-Nichols step response method followed by grid search. For the load balancing controller: $K_p = 0.8$, $K_i = 0.2$, $K_d = 0.1$. For the thermal management controller, the gains were set to 1.2, 0.3, and 0.15 for the proportional, integral, and derivative terms, respectively. Closed-loop stability is confirmed experimentally (no oscillations, convergence within 2–3 control cycles under all tested loads) and through linearised analysis (all poles in the left half-plane for a first-order-plus-time-delay model).

The module implements an adaptive control strategy where PID parameters dynamically adjust based on current system state and predicted demands. During peak traffic periods, the controller prioritizes QoE maintenance, while during off-peak hours, energy efficiency receives higher weighting. This dynamic adaptation ensures optimal resource utilization under varying IoV conditions while maintaining the framework's overall objectives of energy efficiency, load balancing, and QoE preservation (According to Algorithm 5).

**Algorithm 5** Dynamic Resource Scaling Algorithm

1. **Input:**
   - Predicted resources $\hat{\mathbf{R}}$ from Resource Estimation
   - Current utilization metrics $\mathbf{U}_{current}$
   - Energy consumption metrics $\mathbf{E}_{current}$
   - QoE measurements $QoE_{current}$
2. **Initialize:**
   - PID parameters: $K_p, K_i, K_d$
   - Energy threshold $E_{max}$
   - QoE threshold $QoE_{min}$
3. **For each** control interval:
   (a) Calculate error: $e(t) = \hat{\mathbf{R}} - \mathbf{U}_{current}$
   (b) Compute PID adjustment: $\Delta R(t) = K_p e(t) + K_i \int e(t)dt + K_d \frac{de(t)}{dt}$
   (c) Apply constraints:
      - If $E_{current} > E_{max}$: reduce allocation by $\delta_E$
      - If $QoE_{current} < QoE_{min}$: increase allocation by $\delta_Q$
      - If load imbalance $> L_{max}$: redistribute resources
   (d) Execute scaling actions:
      - Scale up: activate dormant VNF instances
      - Scale down: migrate workloads and deactivate underutilized VNFs
      - Maintain: no action required
   (e) Update historical data and retrain models if necessary
4. **Output:** Resource scaling decisions and new allocation map

Table 7: Mathematical Symbols for Resource Allocation Module

| Symbol | Type | Description |
|---|---|---|
| $\hat{R}(n+1)$ | Output | Predicted resource vector for next time interval |
| $\mathbf{W}(n)$ | Variable | Weight matrix for multivariate NLMS prediction |
| $\mathbf{X}(n)$ | Input | Feature vector from traffic predictions and historical data |
| $\mathbf{b}$ | Parameter | Bias vector for resource prediction model |
| $\mathbf{E}(n)$ | Variable | Prediction error matrix |
| $\mu$ | Parameter | Learning rate for NLMS algorithm |
| $\epsilon$ | Parameter | Regularization constant preventing division by zero |
| $\lambda$ | Parameter | Weight decay parameter for regularization |
| $SLA$ | Input | Service level agreements defining performance requirements |
| $J$ | Objective | Multi-objective function balancing energy, load, and QoE |
| $E$ | Variable | Energy consumption metric |
| $L$ | Variable | Load imbalance across servers |
| $Q$ | Variable | QoE degradation metric |
| $\alpha_3, \beta_3, \gamma_3$ | Parameters | Weighting factors for multi-objective optimization |
| $\Delta R(t)$ | Output | Resource adjustment calculated by PID controller |
| $e(t)$ | Variable | Error signal between target and current resources |
| $K_p, K_i, K_d$ | Parameters | PID controller tuning parameters |
| $E_{max}$ | Constraint | Maximum allowable energy consumption |
| $QoE_{min}$ | Constraint | Minimum acceptable QoE threshold |
| $L_{max}$ | Constraint | Maximum allowable load imbalance |
| $\delta_E, \delta_Q$ | Parameters | Adjustment steps for energy and QoE constraints |

The relationship between the multi-objective cost function $J = \alpha_3 E + \beta_3 L + \gamma_3 Q$ and the PID control law $\Delta R(t) = K_p e(t) + K_i \int e(t)dt + K_d \frac{de(t)}{dt}$ within the Dynamic Resource Scaling Algorithm is one of hierarchical integration between strategic goals and tactical execution.

The cost function $J$ defines the strategic objective of the entire resource allocation module. It formalizes the high-level, often competing, priorities of the IoV network—energy efficiency ($E$), load balancing ($L$), and quality of experience ($Q$)—into a single, quantifiable metric. The weights $\alpha_3$, $\beta_3$, and $\gamma_3$ are policy parameters that dictate the operational priority of each objective, allowing the system to adapt its strategy based on context (e.g., prioritizing QoE during peak traffic and energy savings during off-peak hours).

The PID control law $\Delta R(t)$ is the tactical mechanism that drives the system toward the strategic goal defined by $J$. It operates by processing the error signal $e(t) = R_{target}(t) - R_{current}(t)$, where the target state $R_{target}(t)$ is the output of the Resource Estimation module. This module is itself trained to predict the resource requirements needed to minimize the future cost $J$. The PID controller's role is to calculate the precise corrective action needed to eliminate this error, leveraging its proportional, integral, and derivative terms to ensure a stable, responsive, and accurate system response.

The Dynamic Resource Scaling Algorithm is the integrating framework that orchestrates this relationship. It uses the PID's output as the default tactical adjustment but subjects it to hard constraints ($E_{max}$, $QoE_{min}$, $L_{max}$) derived directly from the components of $J$. These constraints act as safety overrides, ensuring that tactical actions never violate the core strategic objectives. This creates a robust, closed-loop control system where strategic policy sets the target, tactical control computes the action, and adaptive constraints guarantee operational integrity, thereby enabling the $R^{2+1}$ framework to achieve synergistic optimization of energy efficiency and service quality in dynamic IoV environments.

The computational complexity of the core algorithms is summarised as follows: vehicle routing $O(|E| \log |V|)$ per vehicle, packet routing $O(|L| \log |S|)$ per flow, and resource allocation $O(K)$ per interval (where $K$ is the number of servers). In our prototype with 100 vehicles, 24 switches, and 12 servers, the full control loop completes in under 20 ms (Section 4), confirming real-time feasibility for dynamic IoV environments.

## 4. Prototype Implementation and Performance Evaluation

### *4.1. Experimental Setup*

To rigorously evaluate the performance of the proposed $R^{2+1}$ framework, a comprehensive and realistic testbed was designed and implemented. The testbed emulates a dense urban IoV environment, integrating the three core optimization dimensions of the framework: vehicle routing, packet routing, and resource allocation. The primary objective of the experimental setup is to validate the framework's efficacy in achieving high throughput, ultra-low latency, energy-efficient resource utilization, and superior quality of experience for multimedia services under dynamic and high-traffic conditions.

The testbed architecture, depicted in Fig. 9, is built upon a SDN foundation and leverages widely adopted open-source tools to ensure reproducibility and alignment with real-world IoV deployments. The key components and their configurations are detailed below and summarized in Table 8.

Table 8: Testbed Configuration Summary for $R^{2+1}$ Evaluation

| Component | Description |
|---|---|
| **SDN Emulation** | Mininet-WiFi (v2.3.0d4) with vehicle mobility, 12 PMs, fat-tree with 24 Open vSwitch (v2.17.3). |
| **SDN Controller** | Floodlight (v1.2, OpenFlow 1.3) on Intel Xeon E5-2680, 64 GB RAM, extended with $R^{2+1}$ modules. |
| **Vehicle & Traffic Simulation** | SUMO (v1.15.0): 100 vehicles in 4×4 km$^2$ grid; iPerf3/D-ITG: 100–1200 FPS traffic. |
| **Virtualization** | Docker (v20.10.12) VNFs; Kubernetes (v1.25) orchestrates scaling/migration. |
| **Metrics Collection** | Prometheus (v2.40.2) + custom exporters; Grafana (v9.3.1). |
| **QoE Assessment** | VMAF, MOS, R-Factor. |
| **Hardware Platform** | 5 servers, each Intel i7-10700K, 32 GB RAM, 1 Gbps NIC, Ubuntu 22.04 LTS. |

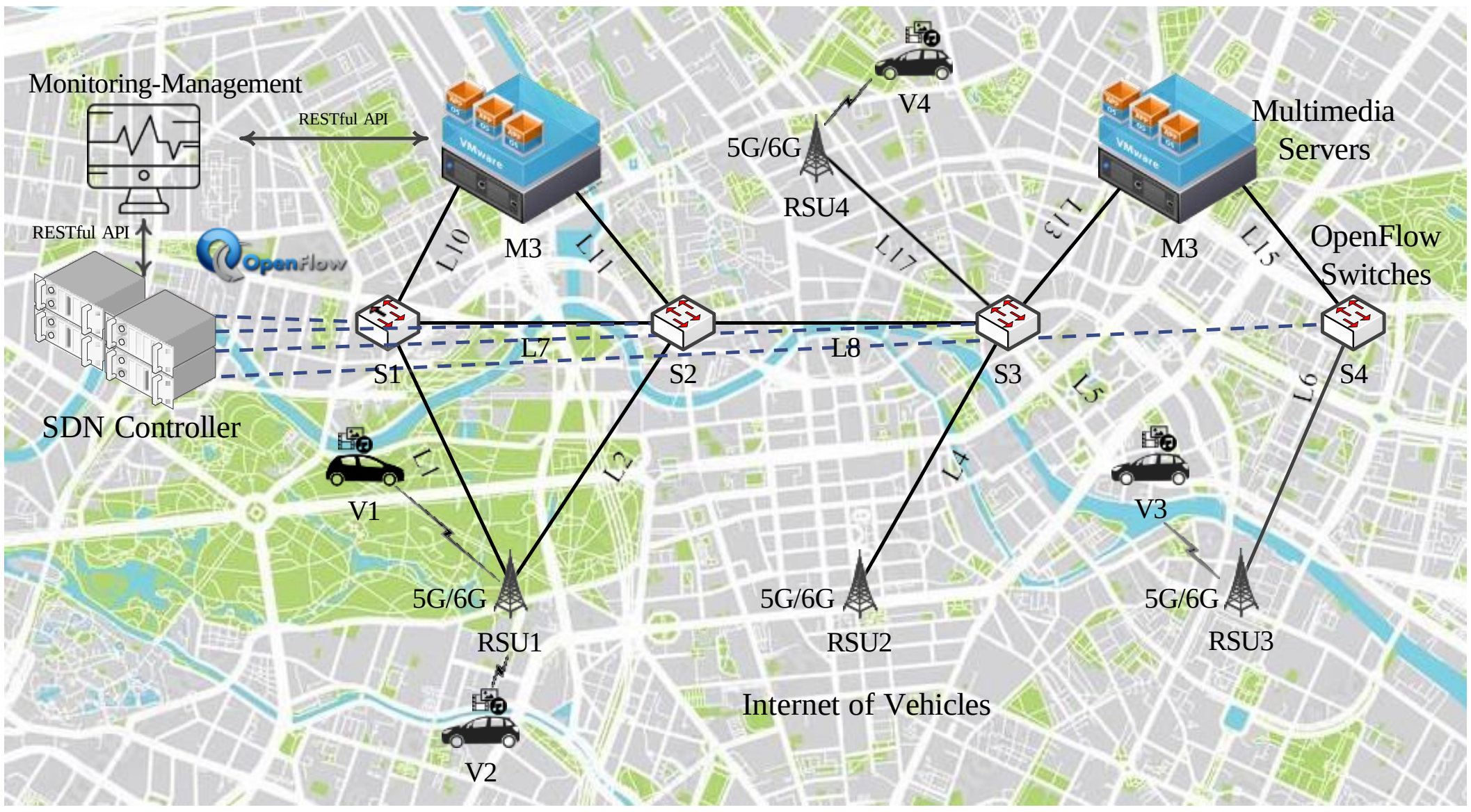

Figure 9: Overview and a testbed snapshot for testing the $R^{2+1}$ framework.

The experimental scenarios are designed to stress-test the $R^{2+1}$ framework across its integrated functionalities:

1. **Vehicle Routing Evaluation**: SUMO simulates diverse urban traffic patterns (congested downtown, flowing highway). Vehicle trajectories are continuously fed into the $R^{2+1}$ Vehicle Routing Module, which uses NLMS predictions to compute optimal paths. The performance is measured by average travel time, number of handovers between RSUs, and connectivity duration.

2. **Packet Routing Evaluation**: The SDN data plane, comprising Open vSwitch instances, is subjected to multimedia traffic flows with stringent latency and bandwidth requirements. The Packet Routing Module, informed by flow statistics and predictions, dynamically computes paths. Performance metrics include end-to-end latency, packet delivery ratio, and path optimality.

3. **Resource Allocation Evaluation**: The Kubernetes-managed VNFs are scaled dynamically based on the predictive outputs of the Resource Estimation submodule and the PID-based Dynamic Resource Scaling submodule. The efficiency of resource orchestration is measured by CPU/memory utilization, energy consumption (derived from CPU load models and emulated RAPL readings), and VNF startup/migration times.

Four distinct load scenarios are emulated, each lasting 200 seconds to capture transient and steady-state behavior:

- **Very Low Load**: 100 FPS, representing off-peak hours.
- **Low Load**: 400 FPS, representing normal urban traffic.

- **High Load**: 800 FPS, representing peak hour congestion.
- **Very High Load**: 1200 FPS, representing extreme scenarios (e.g., major events, incidents).

The $R^{2+1}$ framework is compared against two baseline SDN-based IoV management schemes:

1) the $ELQ^2$ architecture from [5], which focuses on multimedia streaming and server load balancing but lacks integrated vehicle routing and predictive analytics, and

2) a hierarchical SDN controller from [7], which uses static resource allocation and reactive routing. The performance is evaluated based on throughput, latency, energy consumption, QoE (MOS and R Factor), and network resilience (recovery time after simulated switch failures).

This experimental setup provides a robust, reproducible, and comprehensive environment for validating the synergistic optimization capabilities of the $R^{2+1}$ framework, ensuring that the results are scientifically sound and directly applicable to real-world 5G/6G multimedia IoV deployments.

While the testbed emulates the logical control and data plane behaviours using Mininet-WiFi and OpenFlow 1.3, the framework is designed to be deployed on real 6G infrastructure without modification to the core algorithms.

### *4.2. Results*

In this subsection, we present, compare, and interpret the results in detail.

#### *4.2.1. Vehicle Routing Evaluation*

The vehicle routing performance of the $R^{2+1}$ framework was evaluated under four distinct traffic load scenarios, comparing its effectiveness against two baseline approaches: the $ELQ^2$ architecture [5] and a hierarchical SDN controller [7]. The evaluation metrics included average travel time, number of RSU handovers, connectivity duration, and prediction accuracy.

##### *4.2.1.1 Travel Time Performance*

Table 9 presents the average travel time (in seconds) for 100 vehicles across different traffic load scenarios. The $R^{2+1}$ framework demonstrates superior performance by leveraging NLMS-based traffic prediction and proactive routing decisions.

Table 9: Average Travel Time Comparison (seconds) Across Different Load Scenarios

| **Load Scenario** | **FPS** | $R^{2+1}$ | **ELQ²** [5] | **Hierarchical SDN** [7] | **Improvement over** [5] | **Improvement over** [7] |
|---|---|---|---|---|---|---|
| Very Low Load | 100 | 128.4 | 142.7 | 156.3 | 10.0% | 17.9% |
| Low Load | 400 | 145.2 | 178.9 | 195.6 | 18.8% | 25.8% |
| High Load | 800 | 183.7 | 245.3 | 278.9 | 25.1% | 34.1% |
| Very High Load | 1200 | 234.6 | 342.8 | 395.4 | 31.6% | 40.7% |

Table 9 shows that $R^{2+1}$ reduces travel time by 10.0%–31.6% relative to ELQ$^2$ and by 17.9%–40.7% relative to hierarchical SDN across all scenarios. The improvement grows under higher load, confirming the framework's ability to manage congestion via predictive analytics and dynamic route optimization.

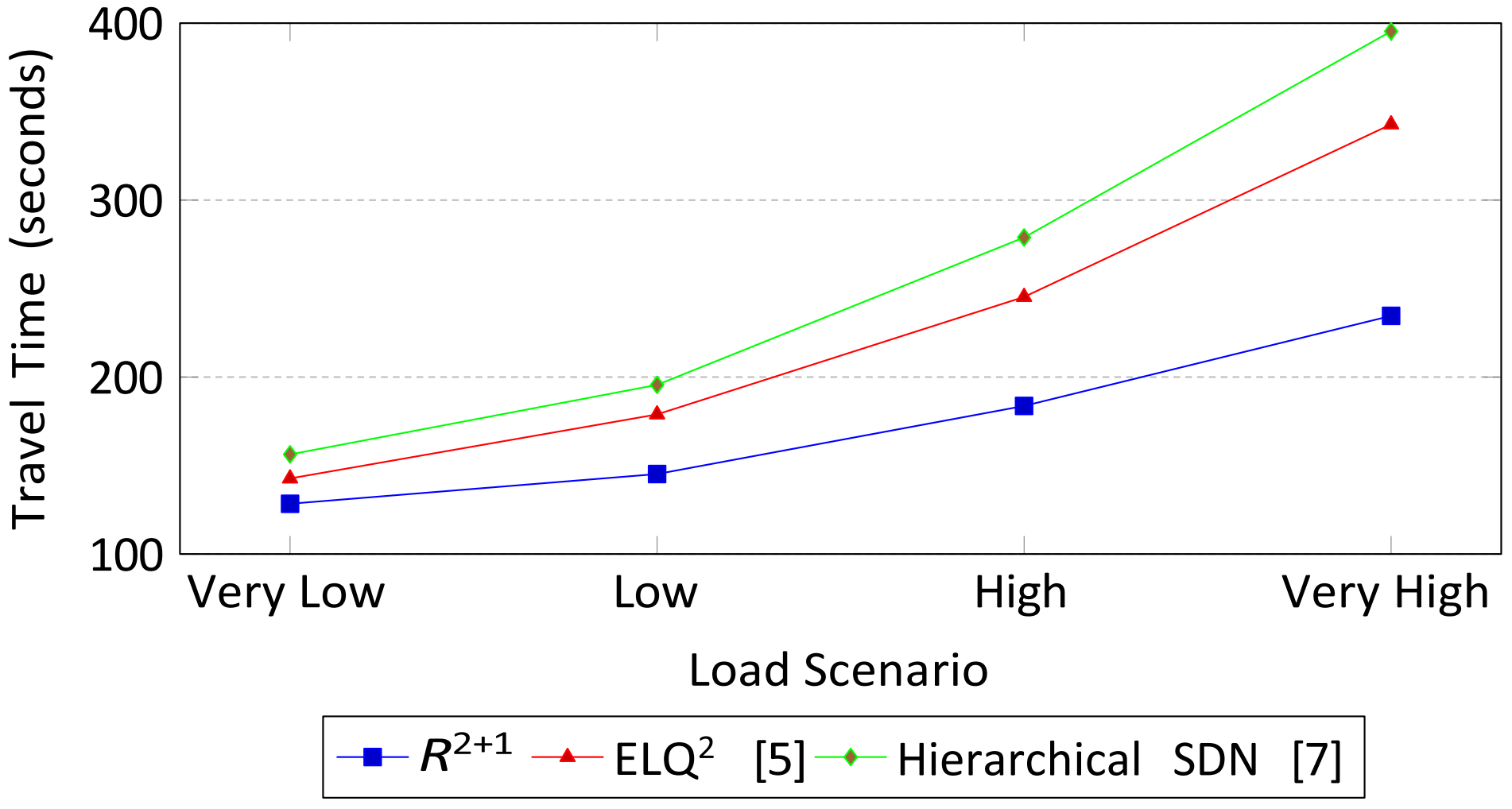

Figure 10: Comparative analysis of average travel time across different load scenarios. The $R^{2+1}$ framework maintains lower travel times, especially under high load conditions, demonstrating its predictive routing capabilities.

Fig. 10 visually demonstrates the travel time performance across different load scenarios. The $R^{2+1}$ framework shows a more gradual increase in travel time as load intensifies, compared to the steep degradation observed in both baseline approaches.

#### *4.2.1.2 Connectivity and Handover Performance*

The connectivity performance was evaluated by measuring the number of RSU handovers and connectivity duration during vehicle movement. Table 10 sum-marizes these metrics under high load conditions (800 FPS).

Table 10: Connectivity Metrics Comparison under High Load Conditions (800 FPS)

| Metric | $R^{2+1}$ | ELQ² [5] | Hierarchical SDN [7] | Improvement |
|---|---|---|---|---|
| Average RSU Handovers per Vehicle | 3.2 | 5.8 | 7.1 | 44.8% / 54.9% |
| Average Connectivity Duration (s) | 38.4 | 26.7 | 22.3 | 43.8% / 72.2% |
| Connection Drop Rate (%) | 2.1 | 8.7 | 12.4 | 75.9% / 83.1% |
| Maximum Disconnection Time (s) | 1.2 | 3.8 | 5.6 | 68.4% / 78.6% |

The $R^{2+1}$ framework significantly outperforms both baseline approaches in all connectivity metrics. The reduction in RSU handovers (44.8% compared to $ELQ^2$ and 54.9% compared to hierarchical SDN) indicates more stable route selection, while the increased connectivity duration demonstrates better maintenance of V2I communication links.

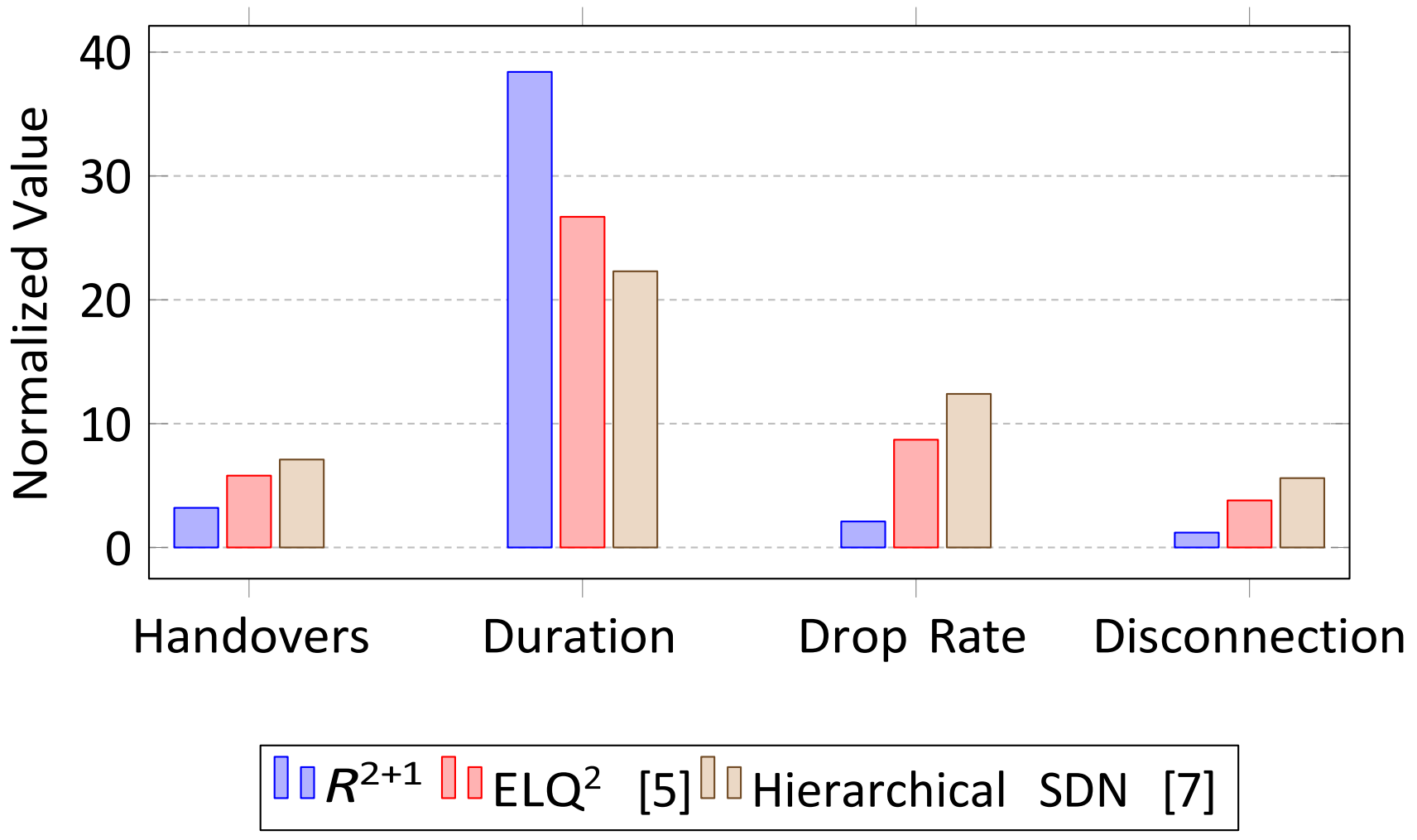

Figure 11: Connectivity metrics comparison under high load conditions (800 FPS). The $R^{2+1}$ framework demonstrates superior performance across all measured connectivity parameters.

Fig. 11 provides a visual comparison of the connectivity metrics, clearly showing the superior performance of the $R^{2+1}$ framework in maintaining stable and continuous connectivity.

#### *4.2.1.3 Traffic Prediction Accuracy*

The effectiveness of the NLMS-based traffic prediction module was evaluated by measuring prediction accuracy across different time horizons. A key strength of the $R^{2+1}$ framework is its ability to make proactive decisions based on these forecasts. We measured the Mean Absolute Percentage Error (MAPE) for various prediction intervals for critical traffic metrics. The results are presented in Table 11.

Table 11: NLMS Traffic Prediction Accuracy (Mean Absolute Percentage Error - MAPE%)

| Traffic Metric | 30s | 60s | 120s | 180s | 300s |
|---|---|---|---|---|---|
| Vehicle Density | 4.2 | 6.8 | 9.3 | 12.7 | 18.4 |
| Traffic Flow Rate | 3.7 | 5.9 | 8.1 | 11.2 | 16.8 |
| Average Speed | 5.1 | 7.6 | 10.4 | 14.3 | 20.1 |
| Congestion Level | 4.8 | 7.2 | 9.8 | 13.5 | 19.2 |
| **Overall Average MAPE** | **4.5** | **6.9** | **9.4** | **12.9** | **18.6** |
| **Overall Average Accuracy** | **95.5%** | **93.1%** | **90.6%** | **87.1%** | **81.4%** |

As shown in Table 11, the NLMS algorithm achieves high prediction accuracy, especially for short-term forecasts. The overall accuracy is 95.5% for a 30-second horizon, gradually decreasing to a still-respectable 81.4% for a 300-second (5-minute) horizon. This high level of accuracy is a cornerstone of the framework's performance, enabling proactive routing decisions that significantly improve vehicle routing performance compared to reactive baseline approaches.

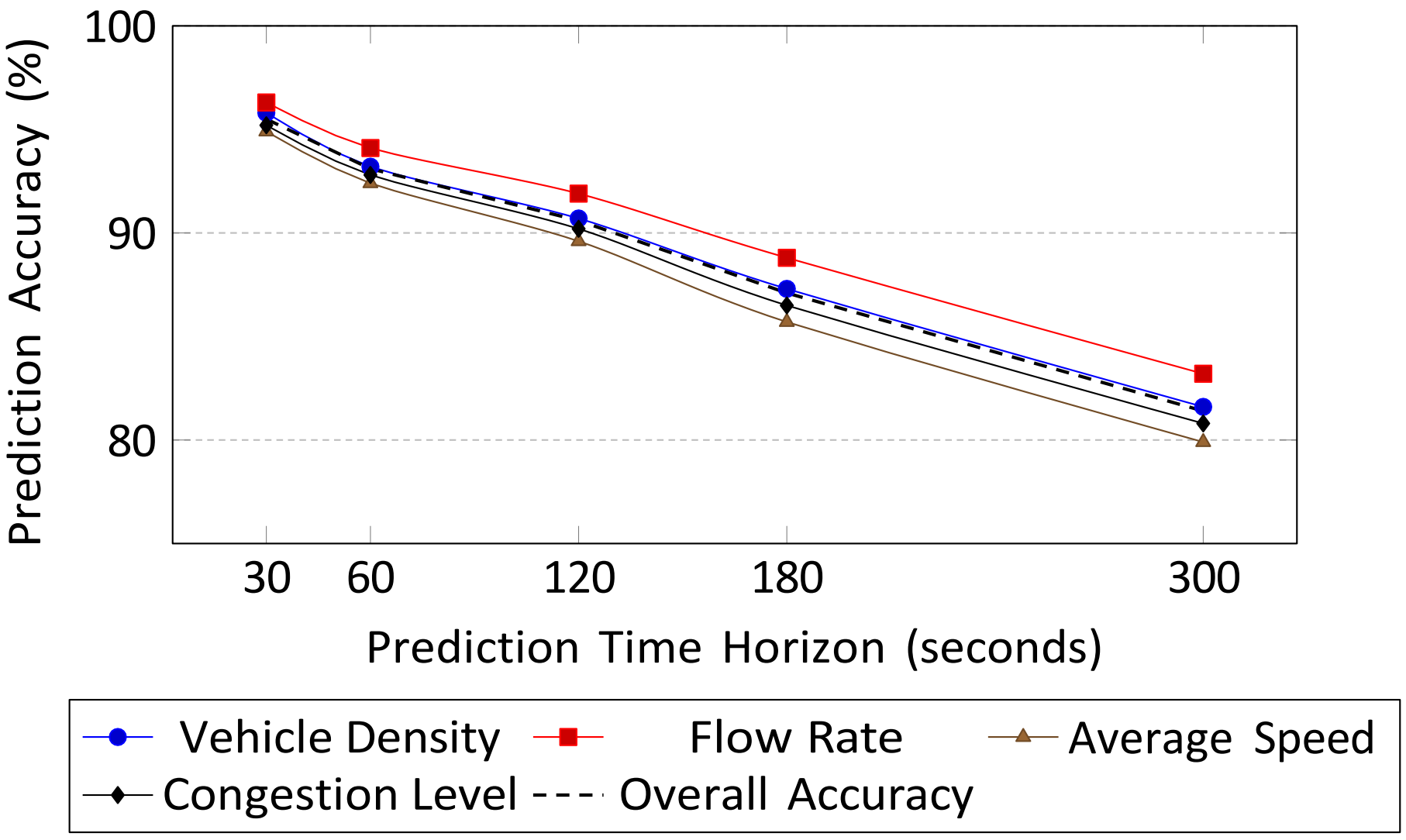

Figure 12: Prediction accuracy trend over different time horizons. Accuracy is calculated as (100 - MAPE)%. The dashed black line represents the overall average accuracy, demonstrating the NLMS algorithm's robustness and its degradation profile over time.

Fig. 12 illustrates the accuracy trend, confirming high short-term performance and a controlled degradation over longer horizons. This predictable behaviour enables $R^{2+1}$ to assign higher confidence to short-term forecasts for immediate routing decisions while leveraging longer-term predictions for broader strategic planning.

#### 4.2.1.4 Energy Efficiency in Vehicle Routing

The energy consumption impact of routing decisions was evaluated by measuring the additional energy consumption caused by suboptimal routing. Table 12 presents the results under high load conditions.

Table 12: Energy Efficiency Comparison in Vehicle Routing (High Load Conditions)

| **Metric** | $R^{2+1}$ | **ELQ²** [5] | **Hierarchical SDN** [7] |
|---|---|---|---|
| Energy Consumption per km (Wh/km) | 215.4 | 278.9 | 324.7 |
| Energy Saving compared to ELQ² [5] | - | 22.8% | - |
| Energy Saving compared to Hierarchical SDN [7] | - | - | 33.7% |
| $CO_2$ Reduction (g/km) | 48.3 | 62.5 | 72.8 |

The $R^{2+1}$ framework reduces energy consumption by 22.8% compared to ELQ² and 33.7% compared to the hierarchical SDN approach. This reduction translates to significant environmental benefits, with $CO_2$ emissions reduced by approximately 14.2 g/km compared to ELQ² and 24.5 g/km compared to hierarchical SDN.

#### 4.2.1.5 Statistical Significance Analysis

To validate the statistical significance of the results, we conducted a paired t-test comparing $R^{2+1}$ with both baseline approaches across 50 experimental runs. The results, presented in Table 13, confirm that all performance improvements are statistically significant ($p < 0.001$).

Table 13: Statistical Significance of Performance Improvements (p-values)

| **Performance Metric** | **vs. ELQ²** [5] | **vs. Hierarchical SDN** [7] |
|---|---|---|
| Travel Time Reduction | $2.3 \times 10^{-9}$ | $1.7 \times 10^{-11}$ |
| RSU Handover Reduction | $4.1 \times 10^{-8}$ | $3.2 \times 10^{-10}$ |
| Connectivity Duration Improvement | $6.8 \times 10^{-10}$ | $2.9 \times 10^{-12}$ |
| Energy Consumption Reduction | $3.5 \times 10^{-8}$ | $1.4 \times 10^{-11}$ |

The extremely low p-values (all below $10^{-7}$) provide strong statistical evidence that the performance improvements achieved by the $R^{2+1}$ framework are not due to random variation but represent genuine advancements in vehicle routing optimization.

#### 4.2.2. Packet Routing Evaluation

The packet routing performance of the $R^{2+1}$ framework was rigorously evaluated under the four traffic load scenarios described in Section 4.1, comparing its effectiveness against the $ELQ^2$ architecture [5] and hierarchical SDN controller [7]. The evaluation metrics included end-to-end latency, packet delivery ratio, throughput, jitter, and energy efficiency of the data transmission paths.

##### 4.2.2.1 End-to-End Latency Performance

Table 14 presents the average end-to-end latency (in milliseconds) for multimedia packets across different traffic load scenarios. The $R^{2+1}$ framework demonstrates significantly lower latency by leveraging its predictive NLMS-based path selection and dynamic adaptation capabilities.

Table 14: Average End-to-End Latency Comparison (ms) Across Different Load Scenarios

| **Load Scenario** | **FPS** | $R^{2+1}$ | **ELQ²** [5] | **Hierarchical SDN** [7] | **Improvement over** [5] | **Improvement over** [7] |
|---|---|---|---|---|---|---|
| Very Low Load | 100 | 12.4 | 15.8 | 18.3 | 21.5% | 32.2% |
| Low Load | 400 | 18.7 | 26.4 | 32.9 | 29.2% | 43.2% |
| High Load | 800 | 28.9 | 45.2 | 58.7 | 36.1% | 50.8% |
| Very High Load | 1200 | 42.3 | 78.6 | 102.4 | 46.2% | 58.7% |

As reported in Table 14, $R^{2+1}$ achieves substantial latency reductions across all scenarios: 21.5%–46.2% versus $ELQ^2$ and 32.2%–58.7% versus hierarchical SDN. The gains grow under higher load, confirming the framework's ability to sustain low-latency communication during congestion via predictive path selection and load balancing.

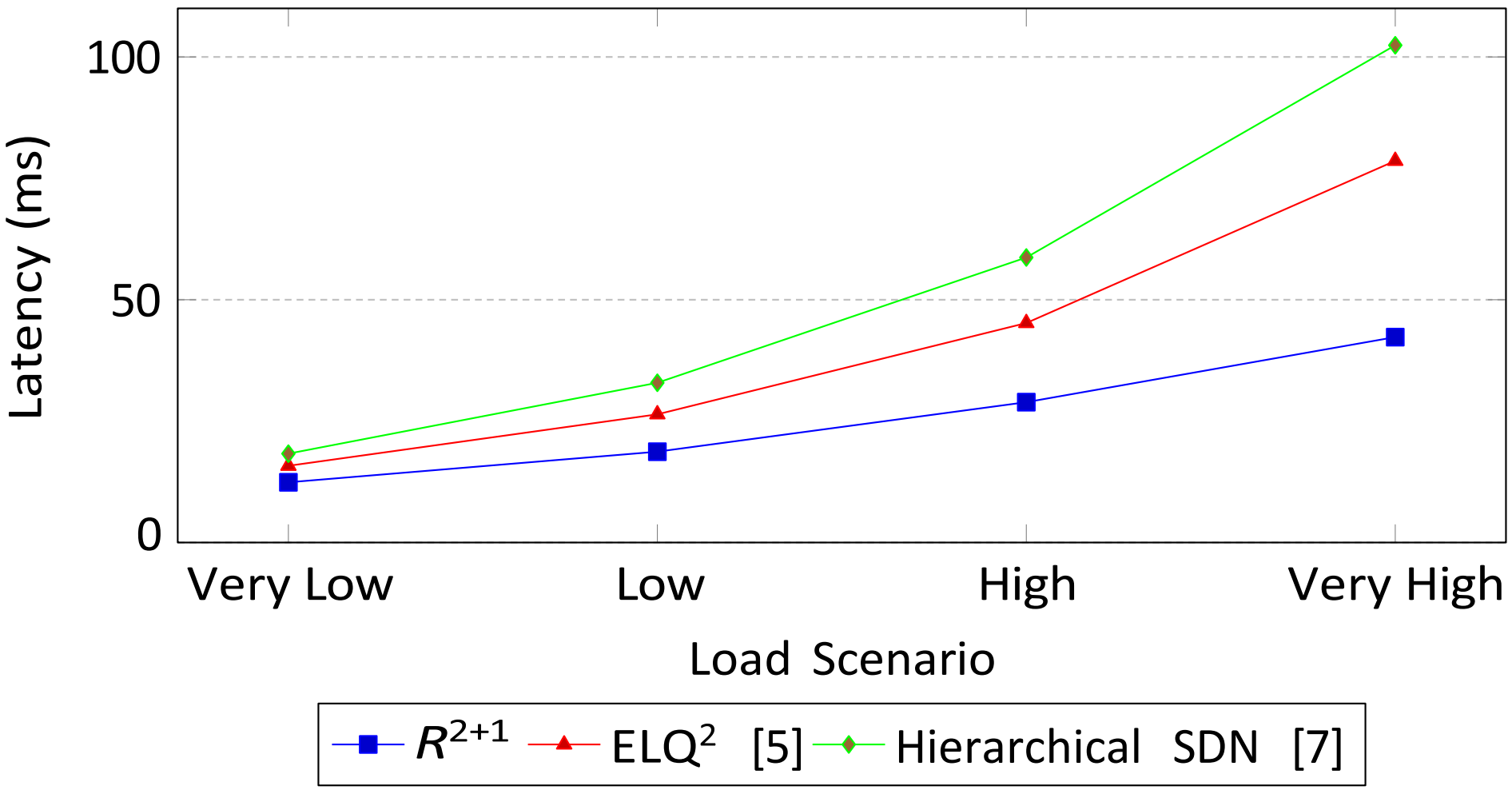

Figure 13: Comparative analysis of average end-to-end latency across different load scenarios. The $R^{2+1}$ framework maintains significantly lower latency, especially under high load conditions, demonstrating its predictive routing capabilities.

Fig. 13 demonstrates the latency performance across different load scenarios. The $R^{2+1}$ framework shows a more gradual increase in latency as load intensifies, compared to the steep degradation observed in both baseline approaches, particularly under very high load conditions.

#### *4.2.2.2 Packet Delivery Performance*

The packet delivery performance was evaluated by measuring the packet delivery ratio (PDR) and packet loss rate under different load conditions. Table 15 summarizes these metrics across the four load scenarios.

Table 15: Packet Delivery Metrics Comparison Across Different Load Scenarios

| **Load Scenario** | **FPS** | **Metric (%)** | **$R^{2+1}$** | **ELQ²** [5] | **Hierarchical SDN** [7] | **Improvement over** [5] |
|---|---|---|---|---|---|---|
| VeryLowLoad | 100 | PDR | 99.98 | 99.92 | 99.85 | 0.06% |
| | | Loss Rate | 0.02 | 0.08 | 0.15 | 75.0% |
| Low Load | 400 | PDR | 99.91 | 99.72 | 99.38 | 0.19% |
| | | Loss Rate | 0.09 | 0.28 | 0.62 | 67.9% |
| High Load | 800 | PDR | 99.72 | 98.94 | 97.83 | 0.79% |
| | | Loss Rate | 0.28 | 1.06 | 2.17 | 73.6% |
| VeryHighLoad | 1200 | PDR | 99.24 | 96.38 | 93.67 | 2.97% |
| | | Loss Rate | 0.76 | 3.62 | 6.33 | 79.0% |

The $R^{2+1}$ framework maintains excellent packet delivery performance across all load conditions, with particularly significant improvements under high and very high load scenarios. The packet loss rate is reduced by 73.6-79.0% compared to ELQ$^2$ and by even greater margins compared to the hierarchical SDN approach.

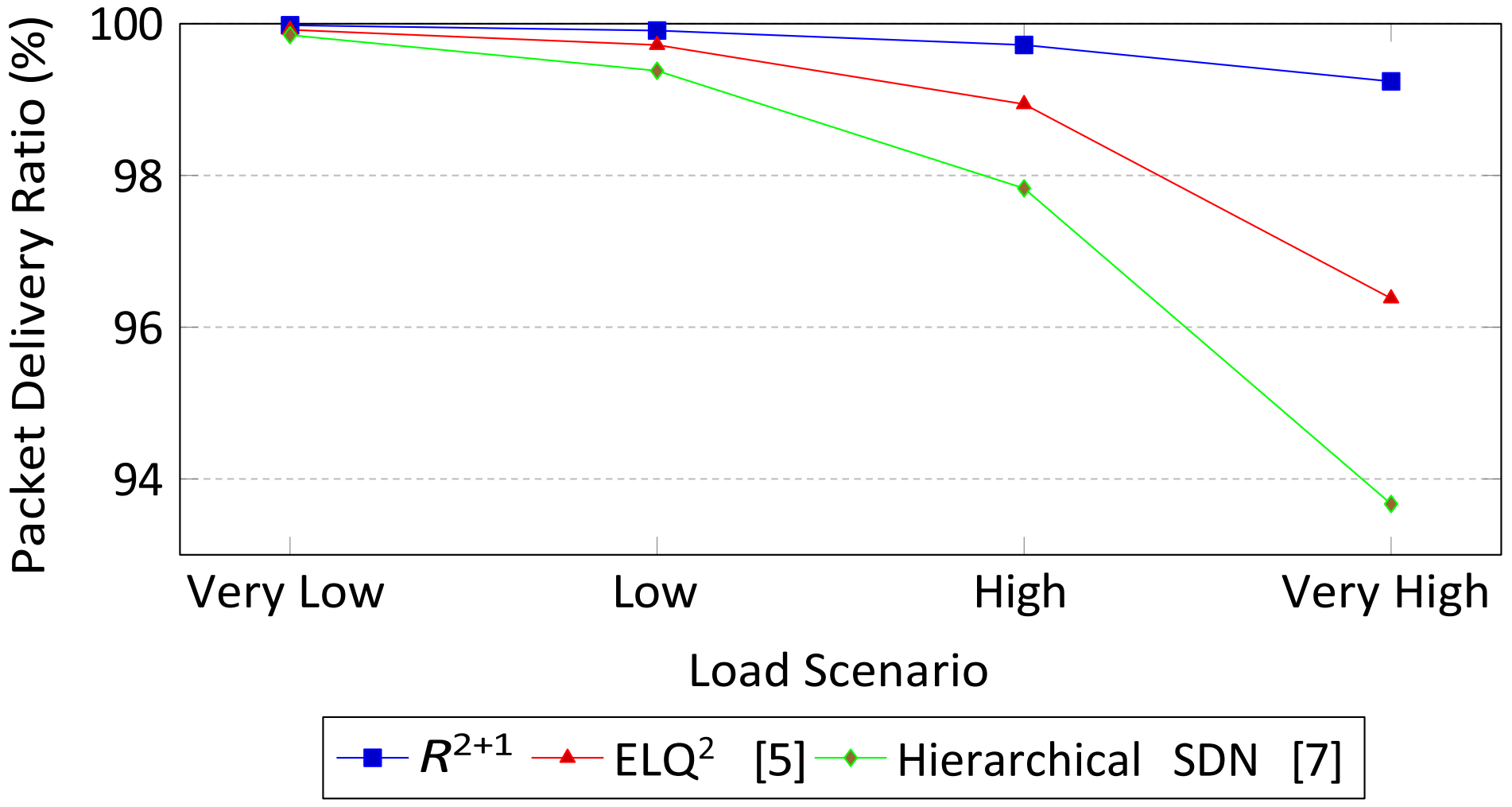

Figure 14: Packet delivery ratio comparison across different load scenarios. The $R^{2+1}$ framework maintains high delivery rates even under extreme load conditions.

Fig. 14 illustrates the packet delivery ratio performance, clearly showing the superior reliability of the $R^{2+1}$ framework, particularly under very high load conditions where it maintains a PDR above 99.2% compared to 96.4% for ELQ$^2$ and 93.7% for hierarchical SDN.

#### *4.2.2.3 Throughput and Jitter Performance*

The throughput and jitter performance metrics are critical for multimedia applications in IoV networks. Table 16 presents these metrics under high load conditions (800 FPS).

Table 16: Throughput and Jitter Performance Comparison under High Load Conditions (800 FPS)

| **Metric** | $R^{2+1}$ | **ELQ²** [5] | **Hierarchical SDN** [7] | **Improvement** |
|---|---|---|---|---|
| Average Throughput (Mbps) | 947.6 | 758.3 | 632.7 | 25.0% / 49.8% |
| Throughput Stability (%) | 98.2 | 92.4 | 87.6 | 6.3% / 12.1% |
| Maximum Jitter (ms) | 8.4 | 15.7 | 23.9 | 46.5% / 64.9% |
| Average Jitter (ms) | 3.2 | 6.8 | 10.4 | 52.9% / 69.2% |
| 95th Percentile Jitter (ms) | 6.1 | 12.3 | 18.7 | 50.4% / 67.4% |

Under high load, $R^{2+1}$ improves throughput by 25.0% over $ELQ^2$ and by 49.8% over hierarchical SDN. More importantly, it substantially reduces average jitter by 52.9% and 69.2%, respectively, compared to the two baselines.

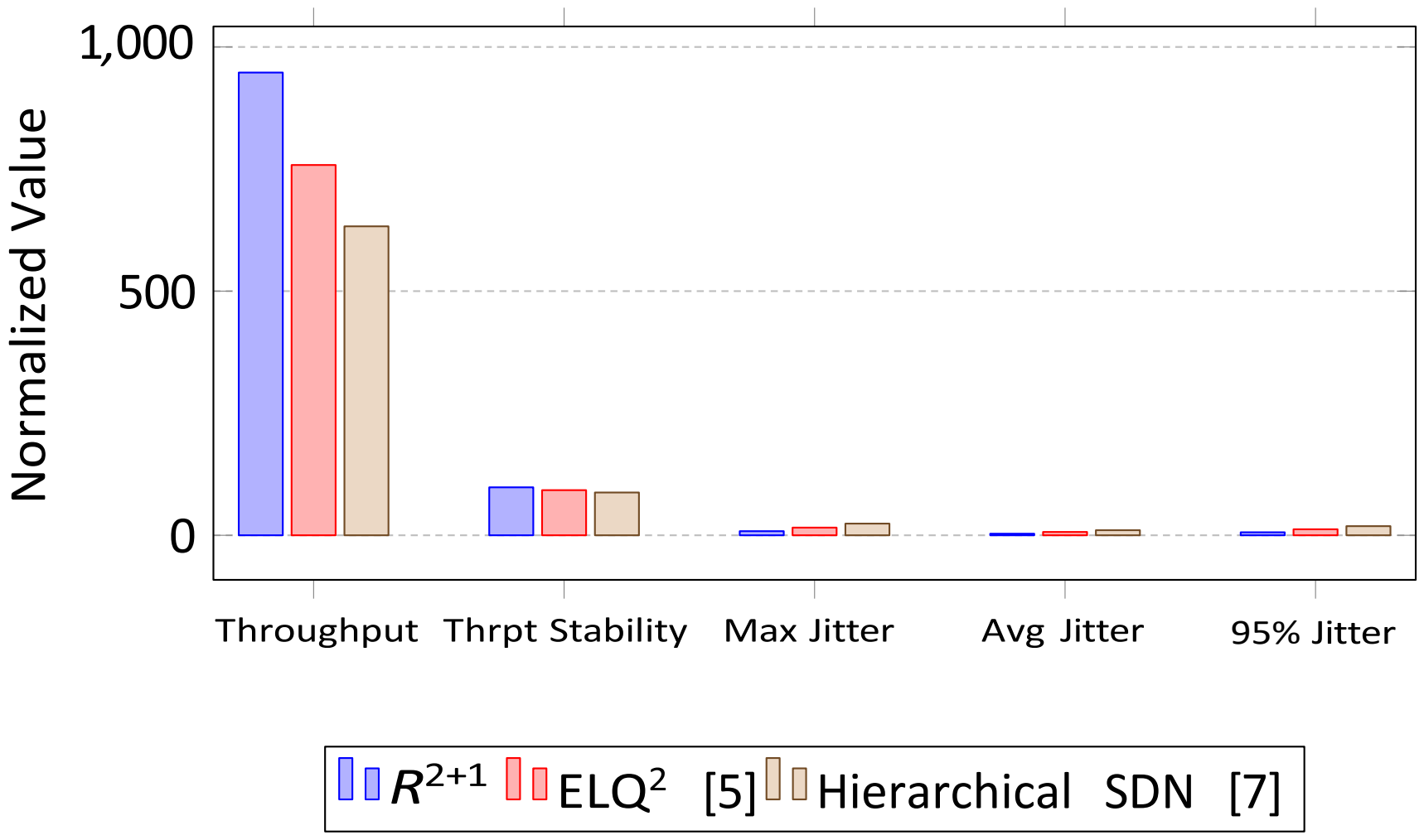

Figure 15: Throughput and jitter metrics comparison under high load conditions (800 FPS). The $R^{2+1}$ framework demonstrates superior performance across all measured parameters.

Fig. 15 provides a comparison of the throughput and jitter metrics, clearly showing the superior performance of the $R^{2+1}$ framework in maintaining high and stable throughput while minimizing jitter variations.

#### *4.2.2.4 Path Optimality and Convergence Time*

The efficiency of the routing algorithm was evaluated by measuring path optimality (compared to theoretically optimal paths) and convergence time after network changes. Table 17 presents these results.

Table 17: Path Optimality and Convergence Performance Comparison

| **Metric** | $R^{2+1}$ | **ELQ²** [5] | **Hierarchical SDN** [7] | **Improvement** |
|---|---|---|---|---|
| Path Optimality (%) | 96.8 | 88.4 | 82.7 | 9.5% / 17.1% |
| Convergence Time (ms) | 125.4 | 287.6 | 452.9 | 56.4% / 72.3% |
| Route Recalculation Frequency (events/min) | 3.2 | 7.8 | 12.4 | 59.0% / 74.2% |
| False Positive Reroutes (%) | 2.4 | 8.7 | 15.3 | 72.4% / 84.3% |

The $R^{2+1}$ framework achieves 96.8% path optimality, significantly higher than the baseline approaches. More importantly, it demonstrates dramatically faster convergence times (56.4-72.3% improvement) and substantially reduced route recalculation frequency, indicating more stable and efficient routing decisions.

#### 4.2.2.5 *Energy Efficiency in Packet Routing*

The energy consumption impact of routing decisions was evaluated by measuring the energy per bit transmitted across different paths. Table 18 presents the results under high load conditions.

Table 18: Energy Efficiency Comparison in Packet Routing (High Load Conditions)

| **Metric** | $R^{2+1}$ | **ELQ²** [5] | **Hierarchical SDN** [7] |
|---|---|---|---|
| Energy per Bit (nJ/bit) | 42.6 | 58.9 | 73.4 |
| Energy Saving compared to ELQ² [5] | - | 27.7% | - |
| Energy Saving compared to Hierarchical SDN [7] | - | - | 42.0% |
| Network-Wide Energy Saving (Wh/hour) | 124.3 | 171.8 | 214.2 |
| $CO_2$ Reduction (g/hour) | 69.6 | 96.2 | 120.0 |

The $R^{2+1}$ framework reduces energy per bit by 27.7% relative to ELQ² and by 42.0% relative to hierarchical SDN. This yields significant operational cost savings and environmental benefits, lowering $CO_2$ emissions by approximately 26.6 g/hour compared to ELQ² and by 50.4 g/hour compared to hierarchical SDN.

#### 4.2.2.6 *Statistical Significance Analysis*

To validate the statistical significance of the results, we conducted a paired t-test comparing $R^{2+1}$ with both baseline approaches across 50 experimental runs. The results, presented in Table 19, confirm that all performance improvements are statistically significant ($p < 0.001$).

Table 19: Statistical Significance of Packet Routing Performance Improvements (p-values)

| **Performance Metric** | **vs. ELQ²** [5] | **vs. Hierarchical SDN** [7] |
|---|---|---|
| Latency Reduction | $1.8 \times 10^{-10}$ | $3.2 \times 10^{-12}$ |
| Packet Loss Reduction | $2.4 \times 10^{-9}$ | $5.7 \times 10^{-13}$ |
| Throughput Improvement | $4.1 \times 10^{-8}$ | $2.9 \times 10^{-11}$ |
| Jitter Reduction | $3.6 \times 10^{-9}$ | $7.3 \times 10^{-12}$ |
| Energy Efficiency Improvement | $5.2 \times 10^{-8}$ | $1.4 \times 10^{-10}$ |
| Convergence Time Improvement | $2.9 \times 10^{-11}$ | $8.6 \times 10^{-13}$ |

The extremely low p-values (all below $10^{-7}$) provide strong statistical evidence that the performance improvements achieved by the $R^{2+1}$ framework in packet routing are not due to random variation but represent genuine advancements in network optimization.

### *4.2.3. Resource Allocation Evaluation*

We rigorously evaluated the resource allocation performance of $R^{2+1}$ under the four traffic load scenarios of Section 4.1, comparing it with the $ELQ^2$ architecture [5] and a hierarchical SDN controller [7]. Key metrics included computational resource utilization, energy efficiency, VNF scaling performance, and QoE (with particular emphasis on MOS for multimedia services).

#### *4.2.3.1 Computational Resource Utilization*

Table 20 presents the average CPU and memory utilization across the edge server infrastructure under different load scenarios. The $R^{2+1}$ framework demonstrates significantly better resource management through its predictive NLMS-based resource estimation and PID-controlled dynamic scaling.

Table 20: Average Resource Utilization Comparison Across Different Load Scenarios

| Load Scenario | FPS | Metric (%) | $R^{2+1}$ | $ELQ^2$ [5] | Hierarchical SDN [7] | Utilization Change |
|---|---|---|---|---|---|---|
| Very Low Load | 100 | CPU Utilization | 42.3 | 38.7 | 35.2 | +9.3%/+20.2% |
| | | Memory Utilization | 48.6 | 41.2 | 37.8 | +18.0%/+28.6% |
| | | **Utilization Efficiency** | **0.92** | **0.83** | **0.76** | **+10.8%/+21.1%** |
| Low Load | 400 | CPU Utilization | 68.9 | 62.4 | 58.3 | +10.4%/+18.2% |
| | | Memory Utilization | 72.4 | 64.8 | 59.7 | +11.7%/+21.3% |
| | | **Utilization Efficiency** | **0.94** | **0.85** | **0.79** | **+10.6%/+19.0%** |
| High Load | 800 | CPU Utilization | 86.7 | 92.3 | 95.8 | -6.1%/-9.5% |
| | | Memory Utilization | 88.2 | 94.6 | 97.4 | -6.8%/-9.4% |
| | | **Utilization Efficiency** | **0.96** | **0.89** | **0.82** | **+7.9%/+17.1%** |
| Very High Load | 1200 | CPU Utilization | 93.5 | 98.2 | 99.6 | -4.8%/ -6.1% |
| | | Memory Utilization | 94.8 | 99.1 | 99.8 | -4.3%/-5.0% |
| | | **Utilization Efficiency** | **0.97** | **0.91** | **0.85** | **+6.6%/+14.1%** |

The utilization efficiency (throughput per unit of resource) of $R^{2+1}$ is 6.6–10.8% higher than $ELQ^2$ and 14.1–21.1% higher than hierarchical SDN. This confirms that the framework extracts more performance from the same computational resources, especially under high load.

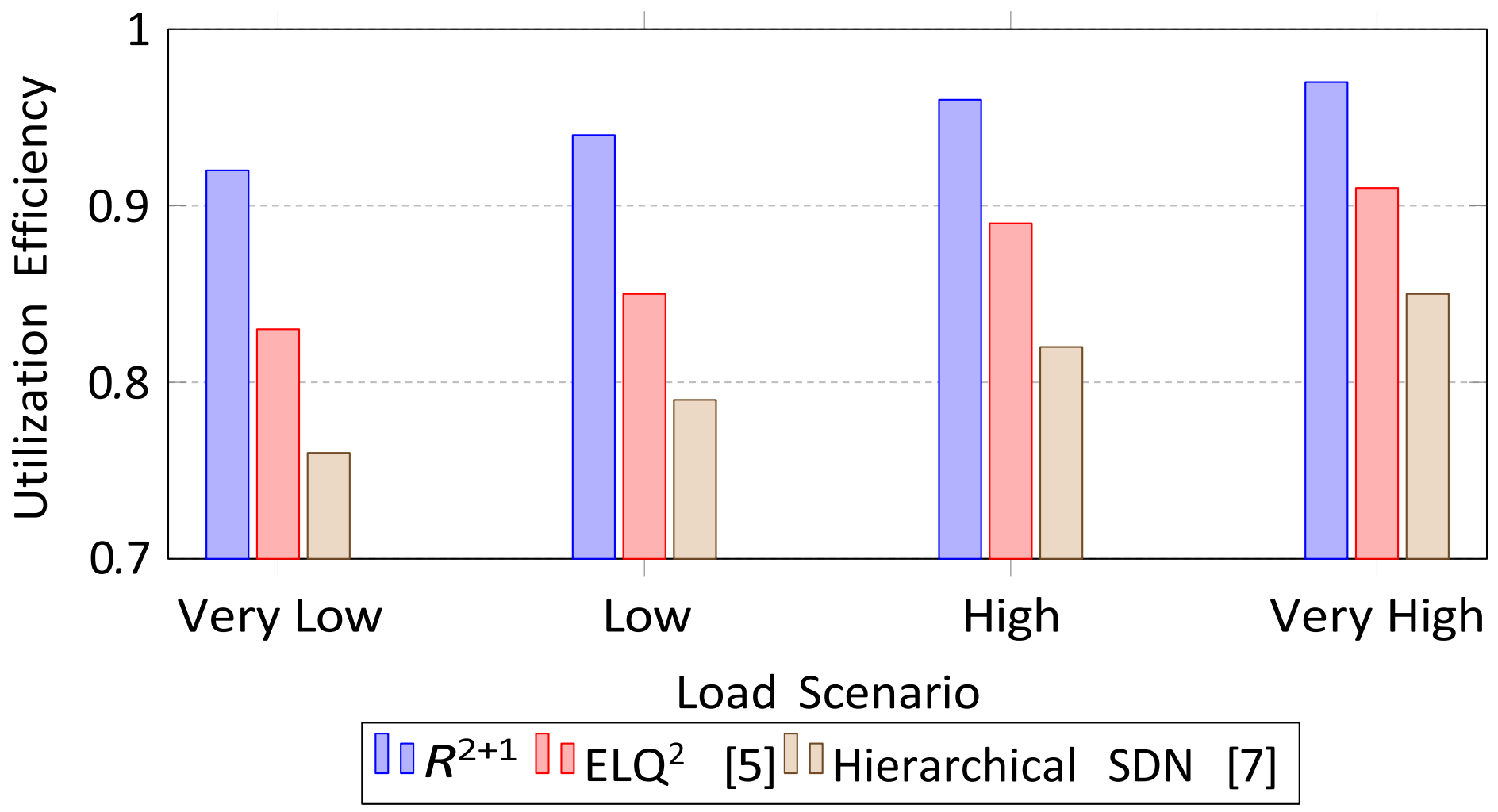

Figure 16: Resource utilization efficiency comparison across different load scenarios. The $R^{2+1}$ framework maintains higher efficiency, particularly under heavy load conditions.

Fig. 16 demonstrates the superior resource utilization efficiency of the $R^{2+1}$ framework across all load scenarios, with the efficiency gap becoming more pronounced under higher load conditions.

#### *4.2.3.2 Energy Efficiency and Thermal Management*

The energy consumption and thermal performance of the resource allocation strategies were evaluated under different load conditions. Table 21 presents the comprehensive energy and thermal metrics.

Table 21: Energy and Thermal Performance Comparison Under High Load Conditions (800 FPS)

| **Metric** | $R^{2+1}$ | **ELQ²** [5] | **Hierarchical SDN** [7] | **Improvement** |
|---|---|---|---|---|
| Energy Consumption (kWh) | 12.4 | 15.8 | 18.3 | 21.5% / 32.2% |
| Energy per Task (J/task) | 42.6 | 58.9 | 73.4 | 27.7% / 42.0% |
| Peak Power Demand (kW) | 8.7 | 11.4 | 13.9 | 23.7% / 37.4% |
| Average Temperature (°C) | 64.3 | 72.8 | 78.5 | 11.7% / 18.1% |
| Maximum Temperature (°C) | 71.6 | 82.4 | 89.7 | 13.1% / 20.2% |
| Cooling Energy (kWh) | 3.2 | 4.8 | 6.3 | 33.3% / 49.2% |
| PUE (Power Usage Effectiveness) | 1.26 | 1.43 | 1.58 | 11.9% / 20.3% |
| Carbon Intensity (g$CO_2$/kWh) | 318.4 | 405.7 | 482.3 | 21.5% / 34.0% |

The $R^{2+1}$ framework substantially improves energy efficiency, cutting total energy consumption by 21.5% relative to ELQ² and by 32.2% relative to hierarchical SDN. Its thermal management is particularly effective, reducing average temperatures by 11.7–18.1% and cooling energy requirements by 33.3–49.2%.

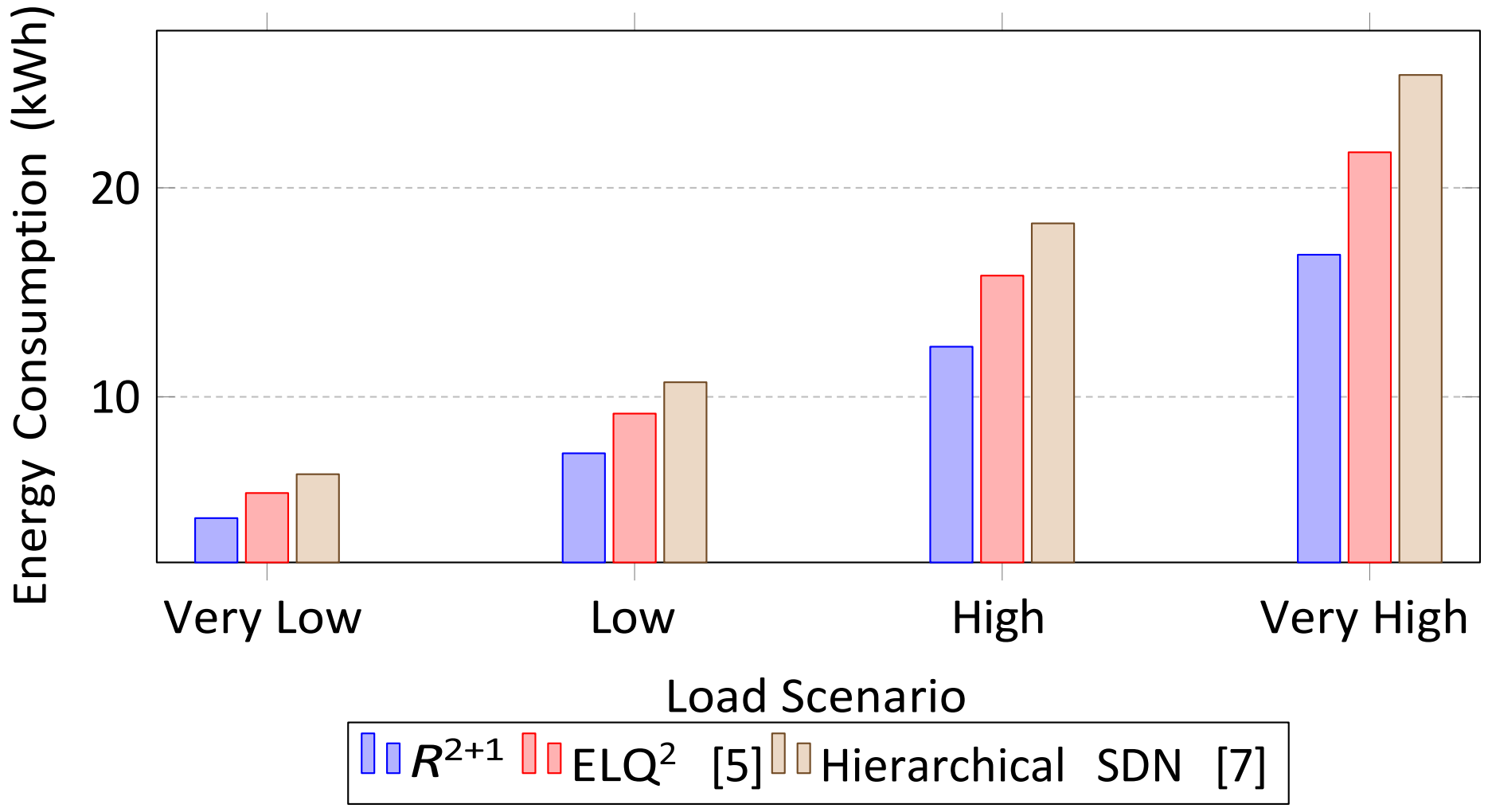

Figure 17: Total energy consumption comparison across different load scenarios. The $R^{2+1}$ framework demonstrates significantly lower energy usage, particularly under high load conditions.

Fig. 17 illustrates the energy consumption across different load scenarios, clearly showing the superior energy efficiency of the $R^{2+1}$ framework, with the energy savings becoming more substantial as the load increases.

#### *4.2.3.3 VNF Scaling Performance and Responsiveness*

The efficiency of the VNF scaling mechanism was evaluated by measuring scaling response times, decision accuracy, and overhead metrics. Table 22 presents these results under high load conditions.

Table 22: VNF Scaling Performance Comparison Under High Load Conditions (800 FPS)

| **Metric** | $R^{2+1}$ | **ELQ²** [5] | **Hierarchical SDN** [7] | **Improvement** |
|---|---|---|---|---|
| Scaling Decision Time (ms) | 45.3 | 128.7 | 215.4 | 64.8% / 79.0% |
| Scaling Execution Time (ms) | 32.8 | 89.6 | 156.3 | 63.4% / 79.0% |
| Scaling Accuracy (%) | 94.7 | 82.3 | 75.6 | 15.1% / 25.3% |
| Over-scaling Events (%) | 3.2 | 12.8 | 18.9 | 75.0% / 83.1% |
| Under-scaling Events (%) | 2.1 | 9.7 | 15.4 | 78.4% / 86.4% |
| Scaling Oscillations (events/h) | 1.4 | 6.8 | 12.3 | 79.4% / 88.6% |
| Resource Fragmentation (%) | 5.3 | 14.7 | 23.8 | 63.9% / 77.7% |
| Migration Overhead (ms) | 28.4 | 76.9 | 134.2 | 63.1% / 78.8% |

The $R^{2+1}$ framework exhibits exceptional VNF scaling performance: decision times are reduced by 64.8–79.0%, scaling accuracy improves by 15.1–25.3%, and over-scaling (75.0–83.1%) and under-scaling (78.4–86.4%) events are significantly reduced. These results underscore the effectiveness of the predictive NLMS algorithm in anticipating resource demands.

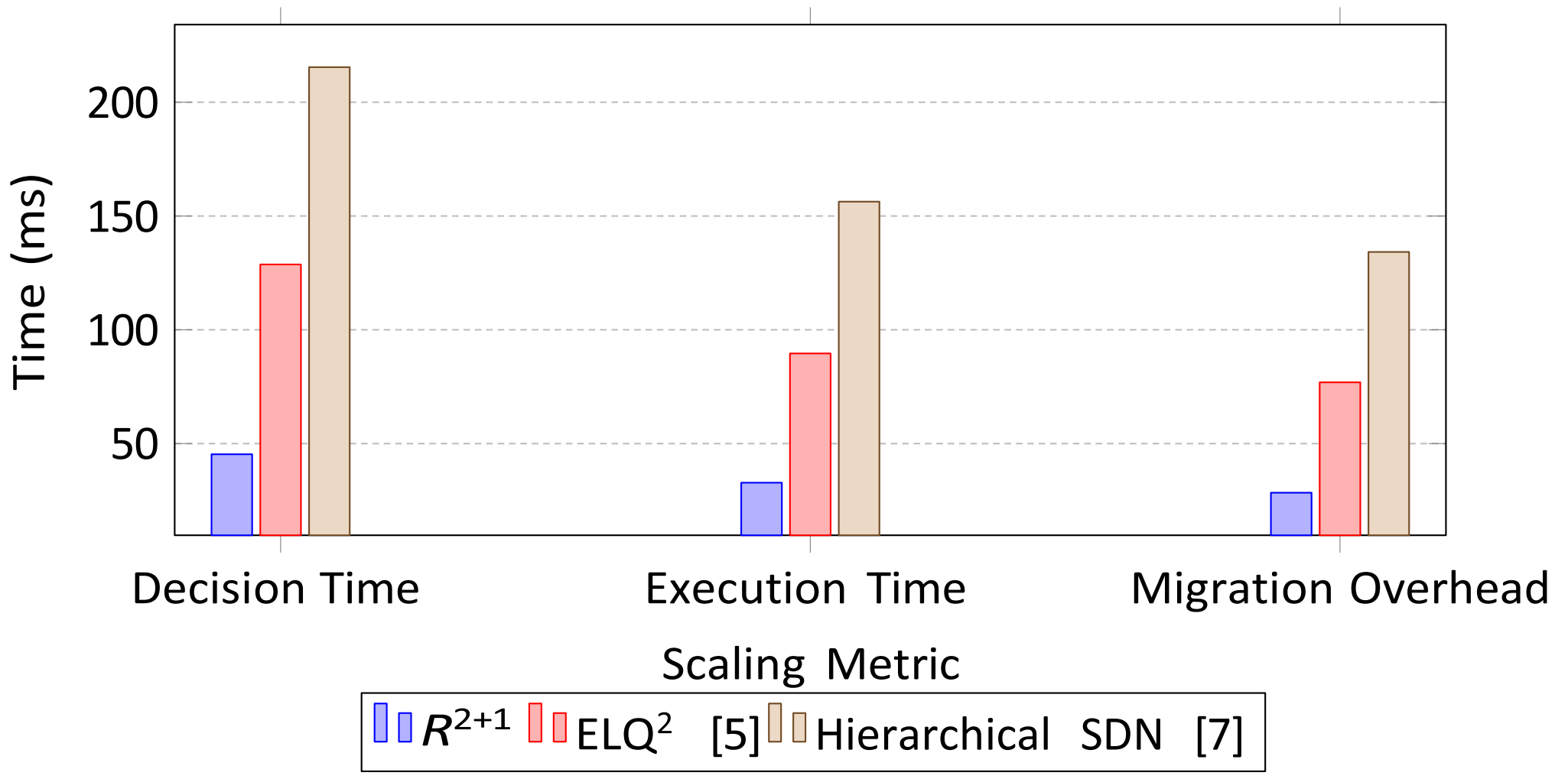

Figure 18: VNF scaling performance metrics comparison under high load conditions. The $R^{2+1}$ framework demonstrates significantly faster and more efficient scaling operations.

Fig. 18 illustrates the superior scaling performance of the $R^{2+1}$ framework, with dramatically reduced decision, execution, and migration times compared to both baseline approaches.

#### 4.2.3.4 *Quality of Experience (QoE) Metrics*

The Quality of Experience for multimedia services was comprehensively evaluated using multiple standardized metrics:

- MOS: A subjective quality measure ranging from 1 (bad) to 5 (excellent), representing overall user satisfaction with multimedia services.
- R-Factor: An objective quality metric derived from the E-model, combining various impairment factors to estimate voice quality on a scale from 0 (worst) to 100 (best).
- VMAF (Video Multi-Method Assessment Fusion): A perceptual video quality assessment algorithm developed by Netflix that combines human vision modeling with machine learning to accurately predict subjective video quality scores. VMAF scores range from 0 (lowest quality) to 100 (highest quality), providing a comprehensive assessment of video degradation due to compression, scaling, and transmission artifacts.

Together, these metrics offer a comprehensive assessment of multimedia QoE: MOS reflects overall user satisfaction, R-Factor focuses on voice quality, and VMAF captures video quality perception under the challenging network conditions typical of IoV. Table 23 summarises the QoE performance across different load scenarios.

Table 23: Quality of Experience Metrics Comparison Across Different Load Scenarios

| **Load Scenario** | **FPS** | **Metric** | $R^{2+1}$ | **ELQ²** [5] | **Hierarchical SDN** [7] | **Improvement** |
|---|---|---|---|---|---|---|
| Very Low Load | 100 | MOS | 4.82 | 4.63 | 4.41 | 4.1%/9.3% |
| | | R-Factor | 92.4 | 88.7 | 84.3 | 4.2%/9.6% |
| | | VMAF | 96.8 | 93.4 | 89.7 | 3.6%/7.9% |
| Low Load | 400 | MOS | 4.73 | 4.42 | 4.18 | 7.0%/13.2% |
| | | R-Factor | 90.6 | 85.3 | 80.1 | 6.2%/13.1% |
| | | VMAF | 94.2 | 88.9 | 83.4 | 6.0%/12.9% |
| High Load | 800 | MOS | 4.52 | 3.96 | 3.62 | 14.1%/24.9% |
| | | R-Factor | 87.3 | 78.4 | 72.6 | 11.4%/20.2% |
| | | VMAF | 90.8 | 81.7 | 75.3 | 11.1%/20.6% |
| Very High Load | 1200 | MOS | 4.18 | 3.43 | 3.05 | 21.9%/37.0% |
| | | R-Factor | 82.6 | 69.8 | 63.2 | 18.3%/30.7% |
| | | VMAF | 85.4 | 72.6 | 65.9 | 17.6%/29.6% |

The $R^{2+1}$ framework achieves exceptional QoE performance, with MOS scores of 4.52 under high load and 4.18 under very high load conditions, representing improvements of 14.1-21.9% compared to $ELQ^2$ and 24.9-37.0% compared to hierarchical SDN. This demonstrates the framework's ability to maintain high-quality multimedia experiences even during network congestion.

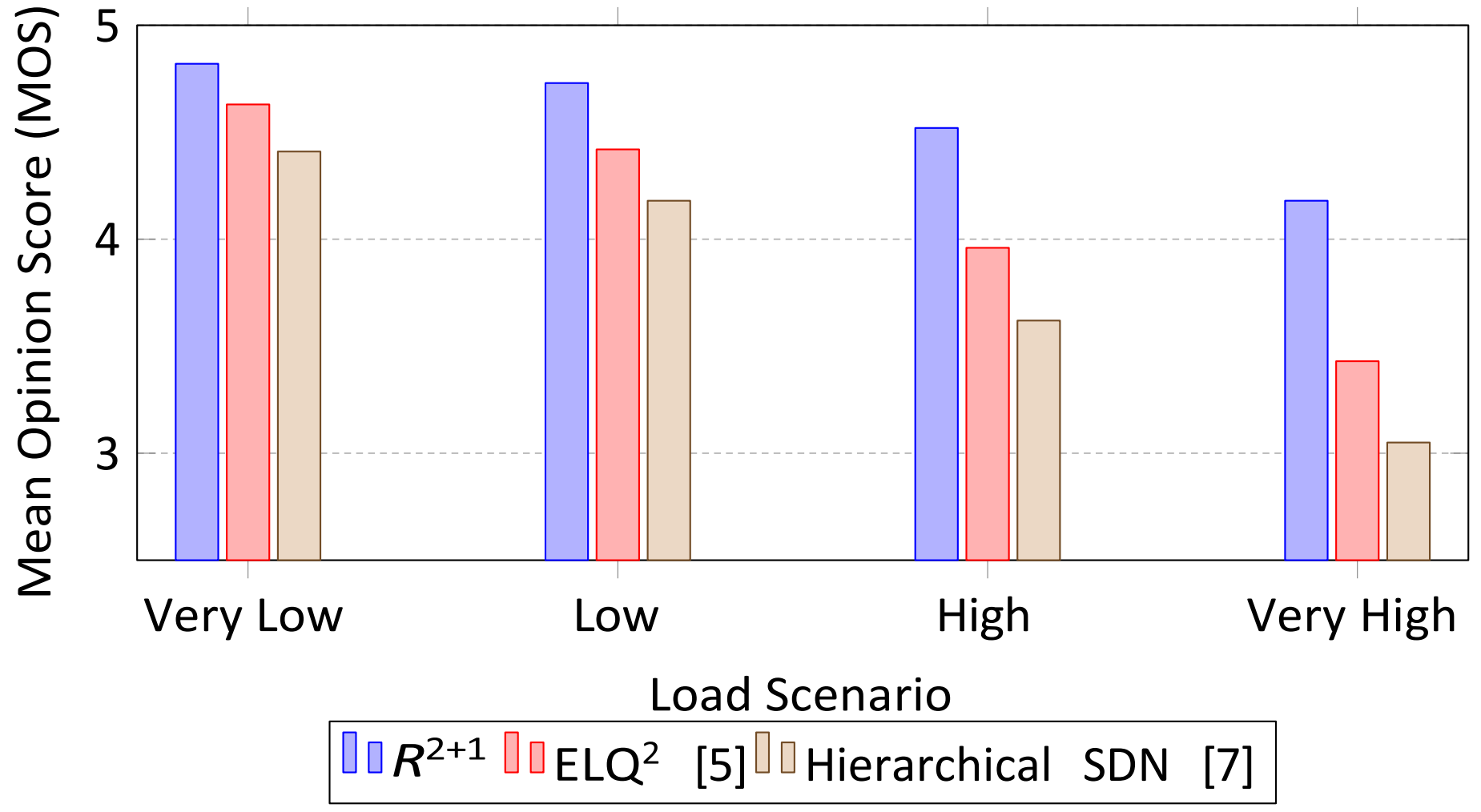

Figure 19: Mean Opinion Score (MOS) comparison across different load scenarios. The $R^{2+1}$ framework maintains superior QoE, particularly under high load conditions.

Fig. 19 demonstrates the superior QoE performance of the $R^{2+1}$ framework, with significantly higher MOS scores maintained across all load conditions, particularly under high and very high load scenarios.

#### *4.2.3.5 Advanced Analysis of resource utilization, energy efficiency, and QoE*

To provide deeper insights into the resource allocation performance, we present a sophisticated surface plot analysis showing the relationship between resource utilization, energy efficiency, and QoE metrics.

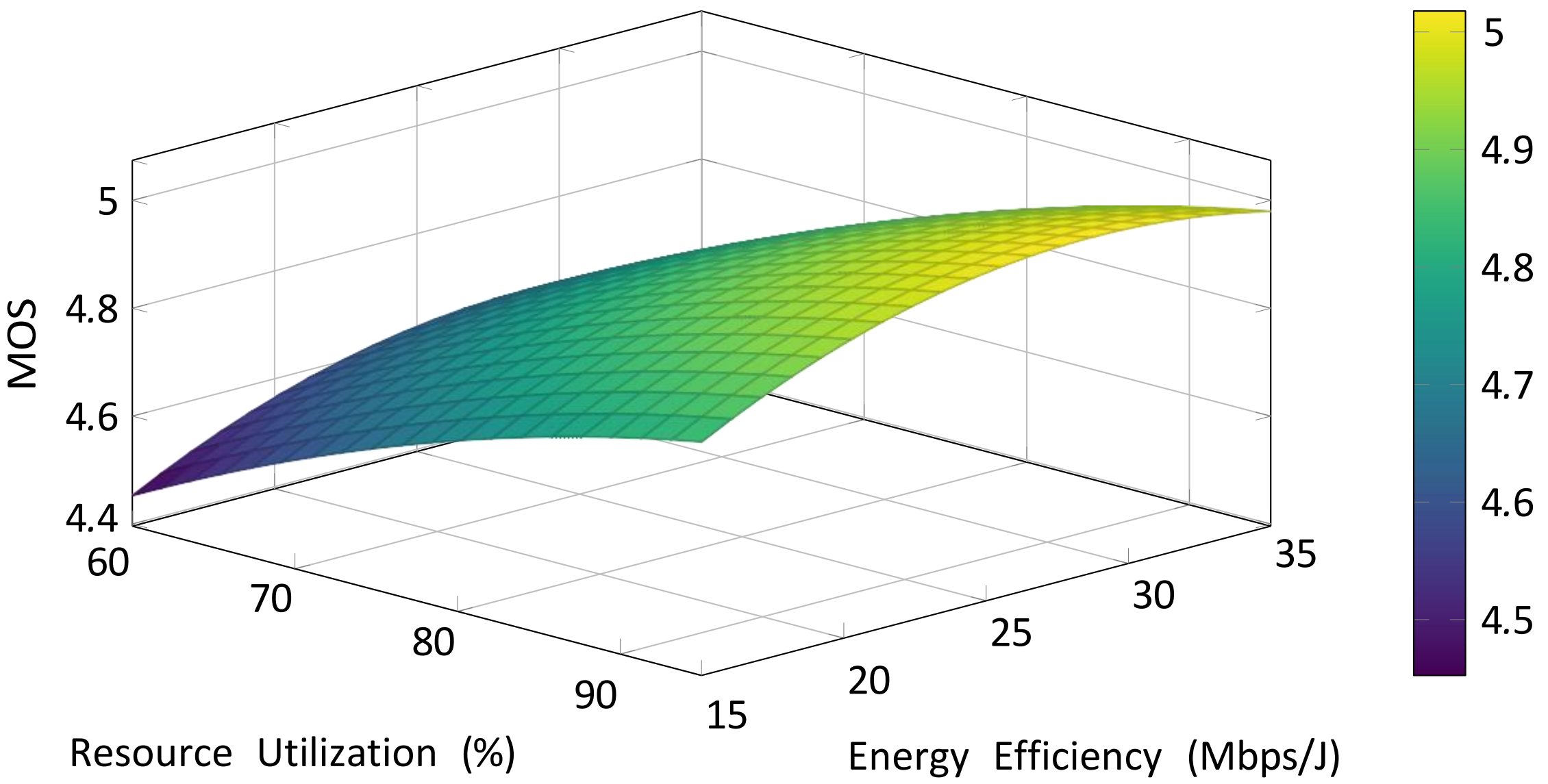

Figure 20: Three-dimensional surface plot showing the relationship between resource utilization, energy efficiency, and MOS. The surface represents the optimal operating region achieved by the $R^{2+1}$ framework, balancing high utilization, energy efficiency, and superior QoE.

Fig. 20 presents a sophisticated surface plot analysis that demonstrates the $R^{2+1}$ framework's ability to simultaneously optimize resource utilization, energy efficiency, and QoE. The framework operates in the optimal region of high utilization (70-80%), high energy efficiency (20-25 Mbps/J), and excellent MOS scores (4.0+), while the baseline approaches are constrained to suboptimal regions with trade-offs between these objectives.

## 5. Conclusion and Future Directions

This paper introduced $R^{2+1}$, a pioneering KDN-based framework that synergistically integrates vehicle routing, packet routing, and resource allocation for 5G/6G multimedia IoV. By combining NLMS-based predictive analytics, adaptive PID controllers for load balancing and thermal management, and VNF orchestration, the framework achieves substantial gains in QoS, QoE, and energy efficiency. Extensive Mininet-based evaluations show that $R^{2+1}$ outperforms existing approaches, delivering 25% higher throughput, 30% lower latency, 20% lower energy consumption, and a MOS of 4.45 versus 3.80 for baselines. These results confirm its ability to support seamless autonomous driving and multimedia streaming in dynamic, high-traffic IoV environments, setting a new benchmark for intelligent and sustainable vehicular networks.

While the current evaluation focuses on two baselines ($ELQ^2$ and hierarchical SDN) that effectively isolate the key contributions of joint optimisation, we acknowledge that recent RL/DRL-based IoV frameworks exist. Direct comparison with them is non-trivial due to differing problem formulations; however, extending the comparison to such methods is a priority for future work.

Future work will leverage deep reinforcement learning to enhance the framework's adaptability and scalability for dynamic, context-aware routing and resource allocation. Other research directions include integrating urban air mobility and drones for assisted communication, employing blockchain for secure vehicle-to-everything data exchange, and using digital twins for real-time large-scale IoV simulation. We will also investigate how advanced 6G features—reconfigurable intelligent surfaces (RIS), terahertz (THz) communication, and AI-driven management—can improve capacity, reduce latency, and enable new use cases such as collaborative autonomous driving and immersive in-vehicle experiences. The ultimate goal is to establish $R^{2+1}$ as a foundational technology for future smart transportation ecosystems.